\documentclass[conference,compsoc,letterpaper]{IEEEtran}
\IEEEoverridecommandlockouts
\usepackage{bbding}
\usepackage{amsmath,amssymb,amsfonts}
\usepackage{amsthm}
\usepackage{mathtools}
\usepackage{bm}

\usepackage{booktabs}
\usepackage{multirow}
\usepackage{array}
\usepackage{subcaption}
\usepackage{enumitem}
\usepackage{tabularx}
\usepackage{makecell}
\usepackage{siunitx}
\usepackage[ruled,linesnumbered,vlined]{algorithm2e}

\usepackage{xspace}
\usepackage{xcolor}
\usepackage{pifont}

\usepackage[hidelinks]{hyperref}
\usepackage[capitalize,noabbrev]{cleveref}

\newtheorem{theorem}{Theorem}[section]
\newtheorem{lemma}[theorem]{Lemma}

\theoremstyle{definition}
\newtheorem{definition}[theorem]{Definition}

\theoremstyle{remark}

\newcommand{\myparagraph}[1]{\noindent \textbf{#1}.}

\newcommand{\myparagraphunderline}[1]{\noindent \underline{#1.}}

\newcommand{\BlindStep}{BlindStep\xspace}
\newcommand{\FBlindStep}{\mathcal{F}_{\textsf{BlindStep}}}
\newcommand{\PiBlindStep}{\Pi_{\textsf{BlindStep}}}
\newcommand{\Leak}{\mathcal{L}}
\newcommand{\LeakSetup}{\Leak_{\mathsf{setup}}}
\newcommand{\LeakExec}{\Leak_{\mathsf{exec}}}
\newcommand{\LeakOut}{\Leak_{\mathsf{out}}}
\newcommand{\LeakAbort}{\Leak_{\mathsf{abort}}}
\newcommand{\Sim}{\mathcal{S}}              %

\newcommand{\share}[1]{[\![#1]\!]}           %
\newcommand{\Falign}{\mathcal{F}_{\mathsf{align}}}
\newcommand{\pp}{\mathsf{pp}}

\newcommand{\Mmax}{M_{\max}}
\newcommand{\compat}{\phi}
\newcommand{\policy}{\rho}
\newcommand{\scorebits}{b}

\newcommand{\FrontierTable}{\mathcal{T}^{F}}
\newcommand{\ExtensionTable}{\mathcal{T}^{E}}
\newcommand{\CandidateTable}{\mathcal{T}^{C}}
\newcommand{\frontierrow}{\mathbf{f}}
\newcommand{\extensionrow}{\mathbf{e}}

\newcommand{\EncTuple}[1]{\tau_{#1}}
\newcommand{\dummytuple}{\tau_{\bot}}
\newcommand{\valid}{\mathsf{v}}
\newcommand{\score}{\mathsf{s}}
\newcommand{\tie}{\mathsf{tie}}
\newcommand{\overflow}{\mathsf{ov}}
\newcommand{\globaloverflow}{\mathsf{ov}_{\mathsf{glob}}}
\newcommand{\PackedTuple}{\mathsf{pt}}

\newcommand{\HSort}{H-Sort}
\newcommand{\HQueue}{H-Queue}
\newcommand{\HSelect}{H-Select}
\newcommand{\BSort}{B-Sort}
\newcommand{\BQueue}{B-Queue}
\newcommand{\BSelect}{B-Select}

\newcommand{\SharedTopK}{\mathsf{TopK}_{\mathsf{shr}}}

\begin{document}

\title{BlindStep: A Secure One-Step Kernel for Privacy-Preserving \\ Collaborative Reasoning}

\author{%
  \IEEEauthorblockN{Xiaoxian Liu\IEEEauthorrefmark{2},
                    Xingyu Tan\Envelope\IEEEauthorrefmark{2}\IEEEauthorrefmark{3}\thanks{\Envelope \ Corresponding author: Xingyu Tan (xingyu.tan@unsw.edu.au).},
                    Chen Chen\IEEEauthorrefmark{4},
                    Xiaoyang Wang\IEEEauthorrefmark{2}}
  \IEEEauthorblockA{\IEEEauthorrefmark{2}University of New South Wales,  Australia, \IEEEauthorrefmark{3}CSIRO, Australia,
  \IEEEauthorrefmark{4}Shenzhen University,  China
  }
  \IEEEauthorblockA{
                    \{xiaoxian.liu, xingyu.tan, xiaoyang.wang1\}@unsw.edu.au,
                    chencc@szu.edu.cn }
}

\maketitle
\begin{abstract}

A recurring step in two-party collaborative workflows combines a private candidate frontier with a partner's compatible evidence, scores the resulting candidates, and retains only a small frontier.
Direct execution can expose intermediate candidates, local scores, or unauthorized evidence provenance.
Existing MPC platforms and PSI protocols provide useful building blocks, but they do not define a unified post-alignment boundary for bounded-slot aggregation, fail-closed overflow handling, retained-frontier selection, and downstream disclosure.
We present BlindStep, a fixed-shape MPC-online kernel for one such post-alignment step between two semi-honest parties.
Given a frontend realizing the bounded-slot contract and an application enforcing the reveal policy $\rho$, BlindStep provides a per-invocation simulation-based guarantee in the $\mathcal F_{\mathsf{align}}$-hybrid model: the kernel leaks only public setup parameters, fixed execution-shape metadata, a global overflow bit, and $\rho$-authorized Layer-2 disclosures.
The guarantee excludes malicious security and multi-invocation composition, and treats frontend leakage and offline preprocessing separately.

Within this kernel boundary, we implement six semi2k realizations in MP-SPDZ, crossing two sharing styles and three top-$K$ strategies: H-variants compare over arithmetic shares, while B-variants convert packed ranking values to Boolean shares before selection.
The Boolean-domain insertion queue \BQueue{} achieves $6$-$7\times$ lower runtime and about $40\times$ lower communication than \HQueue{} at $N=256$ and $N=2048$.
Compared with the Batcher full-ordering reference \HSort{} built on the same semi2k backend, \BQueue{} reduces runtime by $21\times$ and communication by $204\times$ at $N=2048$, and by $30\times$ and $290\times$ at $N=8192$.
Under the evaluated $N=256$, $100$\,ms RTT WAN profile, the fastest realization shifts from \BQueue{} to \HSort{}, showing that round complexity can dominate backend choice for the measured MPC-online kernel.

\end{abstract}

\section{Introduction}
\label{sec:intro}
Collaborative reasoning increasingly requires parties to combine private intermediate state rather than simply exchange final results.
In a typical step, one party holds a private frontier of candidate items, while another party holds evidence that may support or extend those candidates.
After compatible evidence has been aligned, the step must update candidate scores and retain only a small frontier for the next stage.
Directly executing this step can expose the candidate frontier, match pattern, local score contributions, or evidence provenance.
This tension appears in privacy-sensitive retrieval and retrieval-augmented generation~\cite{zeng2024exploring,koga2024privacy,tan2026privgemo}, iterative retrieval~\cite{zai2026proh,sun2019pullnet,sun2024thinkongraph}, knowledge-based reasoning~\cite{tan2025paths,ren2020query2box,tan2025hydrarag,tan2026weaving,tan2026memotime}, and secure collaborative analytics~\cite{liagouris2023secrecy,poddar2021senate,ref3}, with private intermediate inputs.

This paper focuses on this post-alignment step.
We assume that upstream preparation has normalized application objects into typed padded tables and that an alignment frontend has produced bounded secret-shared match slots.
The remaining online kernel must consume these slots, aggregate compatible evidence, enforce the public slot budget without silent truncation, and retain only the top-$K$ frontier.
It should reveal no plaintext beyond a global overflow bit and outputs explicitly authorized by an application-level reveal policy.

Existing techniques provide useful building blocks for this pipeline, but they do not offer a unified post-alignment boundary.
PSI and private-alignment protocols support compatibility discovery, yet they typically leave downstream aggregation, overflow handling, and retained-frontier disclosure to application-specific logic.
SECRECY~\cite{liagouris2023secrecy} and ORQ~\cite{ref3} support collaborative computation, while FSS-based selection~\cite{agarwal2024secure} and three-party secure sorting~\cite{asharov2022efficient} optimize ordering tasks under their own setup and trust assumptions.
What is missing is not another private-alignment protocol or standalone selection primitive, but a narrow post-alignment interface that accounts for bounded-slot consumption, fail-closed overflow handling, top-$K$ retention, and policy-mediated disclosure together for each kernel call.

\begin{figure*}[t]
    \centering
    \includegraphics[width=0.68\linewidth]{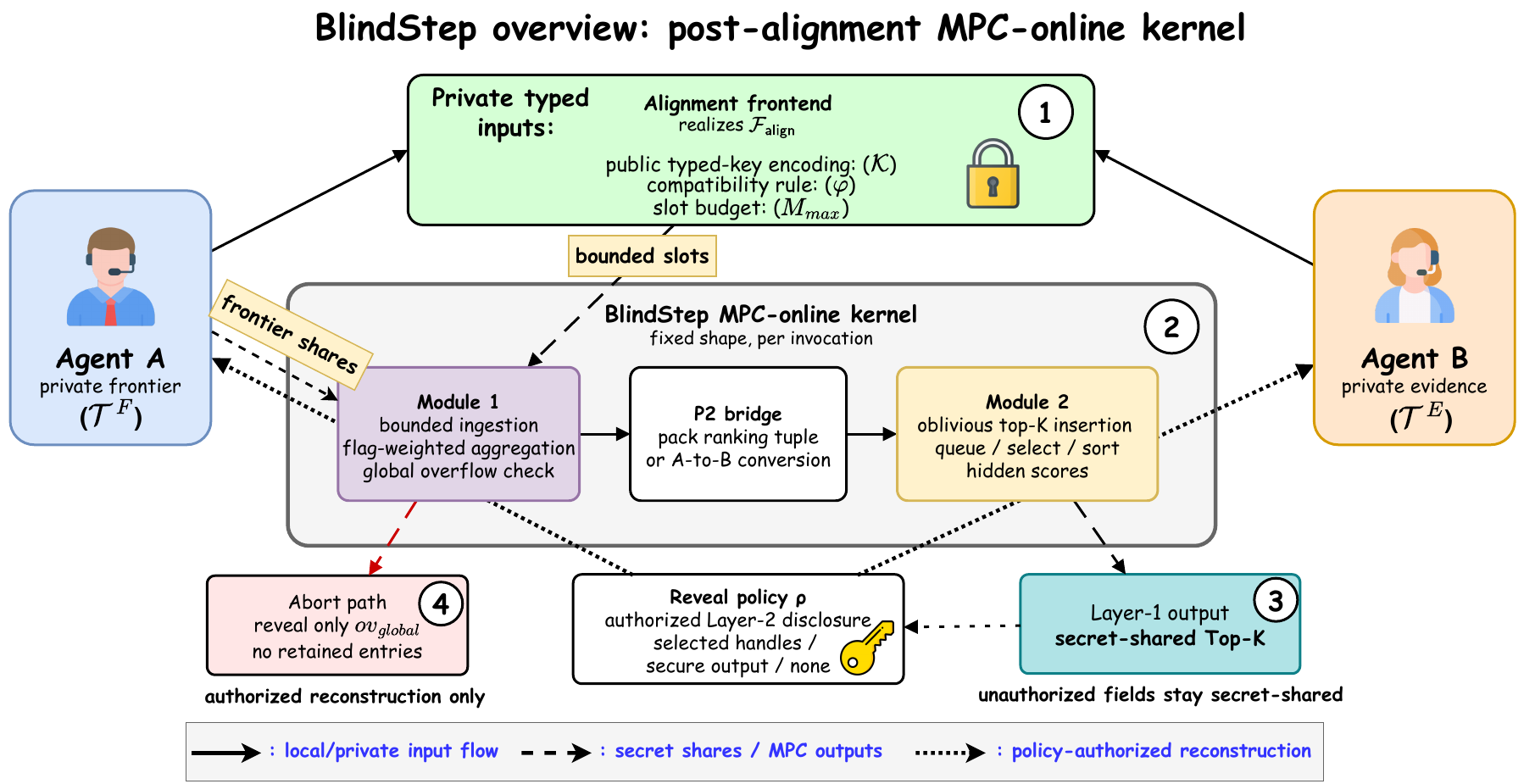}
    \caption{\BlindStep{}  boundary: external alignment supplies shared slots, the kernel aggregates and retains a shared top-$K$ frontier, and external policy $\rho$ governs reconstruction.}
    \label{fig:blindstep_system_boundary}
\end{figure*}

\smallskip
\myparagraph{Contribution}
We present \textbf{\BlindStep{}}, a leakage-bounded building block for the
online MPC phase of two-party collaborative reasoning.
\BlindStep{} gives the recurring post-alignment step a fixed execution
structure and a simulation-checkable leakage boundary.
It is designed to compose with existing alignment frontends, reveal
pipelines, and multi-step orchestration layers rather than replace them.
As shown in Figure~\ref{fig:blindstep_system_boundary}, \BlindStep{} begins
after an alignment frontend has produced bounded secret-shared slots and
ends with a Layer-1 secret-shared retained frontier.
Any plaintext reconstruction is performed outside the kernel according to
the application reveal policy $\rho$.
We model the upstream alignment interface as an abstract functionality
$\Falign{}$. Concrete PSI or Circuit-PSI frontends can instantiate this
interface with their own leakage profiles.
This paper makes the following contributions in interface design and backend evaluation:

\smallskip
\noindent\raisebox{0.2ex}{\scriptsize$\bullet$}\enspace
\textbf{A post-alignment MPC-online kernel with separate leakage accounting.}
We formalize \BlindStep{} as a fixed-shape per-invocation kernel in the
$\Falign{}$-hybrid model, separating frontend alignment leakage,
MPC-online execution leakage, and $\rho$-authorized disclosure leakage.
This separation makes explicit a boundary that general MPC platforms
typically leave to application logic for each invocation.

\smallskip
\noindent\raisebox{0.2ex}{\scriptsize$\bullet$}\enspace
\textbf{A fail-closed bounded-slot interface with workload-grounded checks.}
\BlindStep{} treats $\Mmax$ as a public slot-budget contract over
task-filtered evidence, not a raw-graph multiplicity assumption.
Violations reveal only a global overflow bit, deterministically
aborting before retained entries are released.
Stress tests quantify where simple filters reduce abort risk
and where heavier-tailed workloads need larger budgets or stronger
task-specific evidence pruning.

\smallskip
\noindent\raisebox{0.2ex}{\scriptsize$\bullet$}\enspace
\textbf{Matched-scope backend tradeoffs for the MPC-online kernel.}
We instantiate \BlindStep{} in MP-SPDZ with six MPC-online realizations.\footnote{Artifact: \url{github.com/blindstep-ae/blindstep-artifact} (tag v2.2); archived at \url{doi.org/10.5281/zenodo.23542720}.}
Under LAN, \BQueue{} achieves 6--7$\times$ lower runtime and about
40$\times$ lower communication than \HQueue{} at $N=256$ and $N=2048$,
against the Batcher full-ordering reference on the same semi2k backend,
the gaps reach 30$\times$ runtime and 290$\times$ communication at
$N=8192$.
Under the evaluated $N{=}256$, 100\,ms RTT WAN profile, \HSort{} becomes
fastest despite communicating over $100\times$ more data than \BQueue{},
indicating that round count, rather than bandwidth, is the binding cost in
that setting.
\section{Related Work}
\label{sec:related}

\BlindStep{} connects private alignment, secure aggregation, and top-$K$ retention through a bounded post-alignment interface, with separate leakage accounting for alignment, online execution, and downstream disclosure under $\rho$.

\smallskip
\noindent\textbf{Secure graph and collaborative analytics.}
Graph analytics studies cohesive subgraphs~\cite{tan2023higher}, clique coverage~\cite{kaiyu1}, and neural attributed community search~\cite{jianwei1}.
Related directions include fairness-aware cores~\cite{tan2023maximum}, structural diversity~\cite{kaiyu2}, graph representations~\cite{zhai2025graph}, and hyper-triangle counting~\cite{yin2024hypertriangles}.
Pre-trained graph transformers support unsupervised community search~\cite{jianwei2}, while hyper-truss decomposition~\cite{yin2026hypertruss} and temporal clique compression~\cite{kaiyu3} study cohesive and temporal patterns.
Secure graph frameworks~\cite{nayak2015graphsc,araki2021secure,koti2024graphiti,yu2025graphace} protect graph computation.
GraphOS~\cite{chamani2023graphos} uses trusted hardware for oblivious graph queries, while MultiCent~\cite{bruggemann2025multicent} protects multilayer centrality computation.
Senate~\cite{poddar2021senate}, SECRECY~\cite{liagouris2023secrecy}, and ORQ~\cite{ref3} support secure analytics.
HOLMES~\cite{chang2023holmes} and Cerebro~\cite{zheng2021cerebro} support secure collaborative learning.
\BlindStep{} targets bounded aggregation and hidden-score retention after alignment, with global-only overflow disclosure for the bounded online kernel.

\smallskip
\noindent\textbf{Private alignment and PSI.}
Private set intersection and related private-alignment protocols identify equality matches without revealing full input sets~\cite{rindal2021vole,kolesnikov2016efficient,mohassel2020fast}.
These protocols can instantiate the upstream $\Falign{}$ interface with their own leakage profiles.
The \BlindStep{} kernel consumes its fixed-shape shares of bounded slots, match flags, extension scores, and row-overflow indicators for downstream aggregation and retention of its secret-shared frontier.

\smallskip
\noindent\textbf{Hybrid MPC frameworks.}
MPC frameworks such as MP-SPDZ and ABY combine arithmetic and Boolean representations for different cost profiles~\cite{keller2020mp,demmler2015aby}, with mixed-domain primitives such as daBits and edaBits supporting conversions~\cite{escudero2020improved} for secure mixed-domain computation.

\smallskip
\noindent\textbf{Oblivious sorting and secure top-$K$.}
Oblivious sorting networks~\cite{batcher1968sorting} and secure sorting, merging, and selection protocols~\cite{hamada2012sorting,falk2021secure,asharov2020bucket} provide standard tools for private ordering and selection.
Recent specialized backends include FSS-based sorting and selection~\cite{agarwal2024secure} and three-party radix sorting~\cite{asharov2022efficient}.
Table~\ref{tab:positioning} summarizes their assumptions,
interfaces, and costs.
\BlindStep{} uses such ordering primitives within its post-alignment interface for top-$K$ retention.

\begin{table}[t]
\centering
\caption{Secure sorting and selection backends.}
\label{tab:positioning}
\footnotesize
\setlength{\tabcolsep}{5pt}
\renewcommand{\arraystretch}{1.08}

\begin{tabularx}{\linewidth}{@{}
    >{\raggedright\arraybackslash}p{0.20\linewidth}
    >{\raggedright\arraybackslash}X@{}}
\toprule

\multicolumn{2}{@{}l@{}}{
    \textbf{FSS~\cite{agarwal2024secure}}
} \\
\addlinespace[2pt]

Trust / setup
& Dealer-assisted 2PC or 3PC, one semi-honest
  corruption, protocol-specific correlations. \\

Task
& Shared values $\to$ sorting or rank selection. \\

Cost
& \emph{Selected protocols:}
  $O(n\ell+n\log n)$ online bits,\newline
  2--3 rounds, $O(n^2\ell)$ PRG calls. \\

\midrule

\multicolumn{2}{@{}l@{}}{
    \textbf{3PC radix~\cite{asharov2022efficient}}
} \\
\addlinespace[2pt]

Trust / setup
& Honest-majority 3PC with semi-honest
  and malicious variants. \\

Task
& Shared keys/payloads $\to$ stable full sort. \\

Cost
& \emph{Semi-honest:}
  $O(n\ell\log n+n\ell_p)$ bits/server,\newline
  rounds linear in $\ell$. \\

\bottomrule
\end{tabularx}

\par\vspace{3pt}
{\raggedright\scriptsize
$n$: number of items,
$\ell$: key width,
$\ell_p$: payload width.\par}
\end{table}

\section{System, Threat, and Problem Definition}
\label{sec:system}

This section fixes the boundary used throughout the paper.
We first specify the two-party one-step setting and the typed padded inputs from which the alignment frontend derives fixed-shape shares.
We then state the adversarial model and leakage boundary, separating frontend alignment leakage and downstream disclosure from the \BlindStep{} MPC-online kernel.
Finally, we define the plaintext semantics and output boundary of the one-step functionality.
Figure~\ref{fig:blindstep_system_boundary} provides an overview of \BlindStep{} and its system boundary.
\subsection{System Model}
\label{subsec:system_model}

\noindent\textbf{Parties and one-step scope.}
\BlindStep{} is executed by two semi-honest parties, Agent~$A$ and Agent~$B$, that hold mutually private inputs for one collaborative step.
Agent~$A$ contributes the frontier-side view, and Agent~$B$ contributes the extension-side view.
The functionality returns a Layer-1 retained frontier for this step only.

\smallskip
\noindent\textbf{Typed padded inputs.}
The application presents fixed-size typed padded tables rather than raw graph objects, documents, or database rows.
These tables are the semantic inputs to the one-step interface: upstream preparation normalizes application objects into typed identifiers and scores, and the alignment frontend derives the fixed-shape shares consumed by the online kernel.
The kernel itself does not receive raw identifiers or pairwise match records.
Agent~$A$ provides Frontier Table $\FrontierTable=\{\frontierrow_i\}_{i=0}^{N_F-1}$, where
\[
\frontierrow_i=(\valid_i^{(A)},h_i^{(A)},k_i^{(A)},
\score_i^{(A)},\mathsf{meta}_i^{(A)}),
\]
and Agent~$B$ provides Extension Table$\ExtensionTable=\{\extensionrow_j\}_{j=0}^{N_E-1}$, where
\[
\extensionrow_j=(\valid_j^{(B)},h_j^{(B)},k_j^{(B)},
\score_j^{(B)},\mathsf{meta}_j^{(B)}).
\]
Here $\valid$ marks real versus padding rows, $h$ is an opaque handle, $k$ is an encoded key, $\score$ is a score contribution, and $\mathsf{meta}$ is optional metadata.
Padding rows have $\valid^{(\cdot)}=0$, fields affecting aggregation or retention are gated by validity and never contribute when the row is padding.

\noindent\textbf{Typed key encoding.}
Before alignment, both parties use a public encoding $\mathsf{enc}_{\mathcal K}:\mathcal T\times\mathcal I\to\mathcal K$ to map local objects to typed identifiers in a public key domain
\(\mathcal K\), where \(\mathcal T\) is the public type space and
\(\mathcal I\) is the identifier universe. The spaces
\(\mathcal T,\mathcal I\), encoding format, and output length are public,
concrete local identifiers and encoded keys remain private inputs to the
alignment frontend.

\noindent\textbf{Compatibility rule.}
A public rule $\compat:\mathcal K\times\mathcal K\to\{0,1\}$ defines row compatibility. Concrete keys and the pairwise match pattern remain private except for leakage allowed by the selected alignment frontend and by $\policy$-authorized outputs.
Our artifact compares externally encoded opaque identifiers, $\compat(k_i^{(A)},k_j^{(B)})=\mathbf 1[k_i^{(A)}=k_j^{(B)}]$.
Richer predicates require a different frontend contract and leakage analysis.

\noindent\textbf{Alignment frontend contract.}
The \BlindStep{} security analysis is relative to a chosen realization of $\Falign{}$.
After applying the public compatibility rule, the online kernel starts from Agent~$A$'s frontier-side shares together with fixed-shape $\Falign{}$ outputs: bounded slots, row-overflow shares, match-flag shares, and extension-side score shares.
Any frontend-specific leakage, trust assumption, or input-consistency requirement composes with the kernel leakage $\Leak$, these properties are external to the \BlindStep{} MPC-online analysis and are not re-proved here.

\noindent\textbf{Bounded-slot contract.}
The public slot budget $\Mmax$ is a deployment contract over upstream-filtered evidence, not a raw-graph degree assumption.
Deployments must validate this bound after task-specific filtering or evidence pre-selection.
If a frontier row has more than $\Mmax$ compatible extension slots, the frontend sets the corresponding row-overflow share.
\BlindStep{} then opens only the global overflow bit and aborts fail-closed as in Section~\ref{subsec:module1}.

\noindent\textbf{Public parameters and output policy.}
For presentation and evaluation, we use the symmetric setting $N_F=N_E=N_C=N$, where $N_C$ is the padded candidate-table size consumed by the online kernel, and write $\pp=(N,K,\scorebits,\Mmax,\compat,\policy)$.
Here $N$ is the padded candidate-table size used by the online kernel in the symmetric setting, $K$ is the retention target, $\scorebits$ is the aggregate score width, $\Mmax$ is the slot budget, $\compat$ is the compatibility rule, and $\policy$ is the reveal policy.
We use $b_{\mathsf{in}}$ for the input score width.
The public key domain and encoding convention are part of the selected frontend setup and the type signature of $\compat$, not separate per-invocation kernel parameters.
Asymmetric table sizes are supported: $N_E$ affects frontend cost, while the online kernel scales with $N_C$ and $\Mmax$.

\noindent\textbf{Input preparation and scope.}
\BlindStep{} assumes, but does not implement, upstream preparation that applies the public key encoding and normalizes scores to $[0,2^{b_{\mathsf{in}}})$.
The aggregate width $\scorebits$ must hold one frontier-side contribution and up to $\Mmax$ extension-side contributions.
Raw graph extraction, entity resolution, schema alignment, score normalization, richer predicate compilation, and multi-step orchestration are outside the \BlindStep{} boundary.
Tool-schema planning~\cite{zhai2026hyperagent} and contract-preserving skill reuse~\cite{tan2026skillzip} are examples of such application-level orchestration.

\subsection{Threat Model and Leakage}
\label{subsec:threat_model}

\noindent\textbf{Adversarial model.}
We consider static semi-honest corruption of at most one party over authenticated private channels~\cite{oded2009foundations,lindell2017simulate}.
Parties follow the protocol but may infer from their inputs, randomness, messages, and authorized outputs.
We do not address malicious deviations, selective aborts, adaptive corruption, or implementation side channels.
The deterministic overflow abort in Section~\ref{subsec:module1} preserves correctness under a public slot-budget violation, it does not address malicious selective-abort attacks.
Table~\ref{tab:tcb} summarizes the trusted components and assumptions used by the kernel theorem.
Two components sit outside the MPC-online kernel: the alignment frontend that realizes $\Falign{}$ and the application mechanism that enforces $\policy$.
Their leakage and correctness obligations are accounted for separately and compose with the kernel leakage formalized in Section~\ref{subsec:leakage}.

\begin{table}[t]
\centering
\small
\setlength{\tabcolsep}{3pt}
\caption{Trusted components and assumptions.}
\label{tab:tcb}
\begin{tabular}{@{}p{0.32\columnwidth}p{0.64\columnwidth}@{}}
\toprule
\textbf{Component} & \textbf{Assumption} \\
\midrule
MP-SPDZ backend &
Semi-honest 2PC correctness and privacy \\
Alignment frontend &
Chosen frontend realizes $\Falign{}$, concrete Circuit-PSI instantiation
is accounted for in Sections~\ref{subsec:leakage} and~\ref{subsec:impl} \\
Preprocessing &
Triples follow backend assumptions, edaBits follow
mixed-domain preprocessing assumptions~\cite{escudero2020improved} \\
Channels &
Authenticated and private \\
Public parameters &
Agreed before execution \\
Reveal policy $\policy$ &
Enforced outside the MPC-online kernel \\
\bottomrule
\end{tabular}
\vspace{-3mm}
\end{table}

\smallskip

\noindent\textbf{Protection goals and leakage boundary.}
Within the MPC-online kernel, \BlindStep{} protects intermediate candidates, aggregate and local score information, evidence attribution, and non-selected entries, subject to the leakage of the selected $\Falign{}$ realization and to $\policy$-authorized outputs.
The kernel reveals only public setup parameters, fixed execution-shape metadata, the global overflow bit $\globaloverflow$, and $\policy$-authorized Layer-2 outputs.
Section~\ref{subsec:leakage} formalizes these components as $\Leak=(\LeakSetup,\LeakExec,\LeakAbort,\LeakOut)$.
\smallskip

\noindent\textbf{Reveal policy.}
The public policy $\policy$ maps Layer-1 retained entries to authorized Layer-2 disclosures, such as selected handles to one or both parties, opaque retained entries to downstream secure computation, or no plaintext output.
It does not authorize non-retained entries, local score contributions, or internal tie metadata by default.
Enforcement is application-side: the kernel emits only Layer-1 secret shares, and the ideal functionality uses $\policy$ only to account for $\LeakOut$.
Deployments must bind $\policy$ to their disclosure pipeline.
\smallskip

\noindent\textbf{Invocation scope.}
The model is per-invocation, larger workflows compose leakage across invocations as discussed in Section~\ref{subsec:repeated_invocation}.

\subsection{Problem Definition}
\label{subsec:problem_definition}

With the system and leakage boundaries fixed, the one-step functionality has two semantic cases.
If the bounded-slot contract is violated, the step aborts before releasing retained entries.
Otherwise, it aggregates compatible evidence into one candidate per frontier row and retains only the top-$K$ candidates.

\begin{definition}[{Plaintext semantics}]
Given bounded-slot outputs from $\Falign{}$ and frontier-side contributions, \BlindStep{} first opens the global overflow bit computed from the row-overflow indicators.
If any row overflows, $\globaloverflow=1$ and the step releases no retained entries.
If no row overflows, it produces one aggregated candidate per frontier row rather than one output per compatible pair.
A candidate is valid if the frontier row is valid and has at least one compatible extension, candidates are ordered lexicographically by $(\valid,\score,-\tie)$, where $\tie_i$ is an independent per-execution tie-breaking tag.
The kernel retains only the top-$K$ rows and does not reveal a global order over all candidates.

\end{definition}

\begin{definition}[{Output boundary}]
On the non-abort path, the kernel outputs a Layer-1 retained set:
$
\SharedTopK=\{\share{\mathsf{entry}_\ell}\}_{\ell=0}^{K-1},
\mathsf{entry}_\ell=(\mathsf{payload}_\ell,\valid_\ell,\score_\ell,\tie_\ell).
$
The payload contains an opaque handle, encoded key, and metadata fields.
All retained fields remain secret-shared at the kernel boundary.
They are opened only as authorized by $\policy$, internal tie metadata remains closed by default.
\end{definition}

\section{The \BlindStep{} Protocol}
\label{sec:protocol}

\subsection{Protocol Overview}
\label{subsec:overview}

\begin{figure}[t]
    \centering
    \includegraphics[width=1\columnwidth]{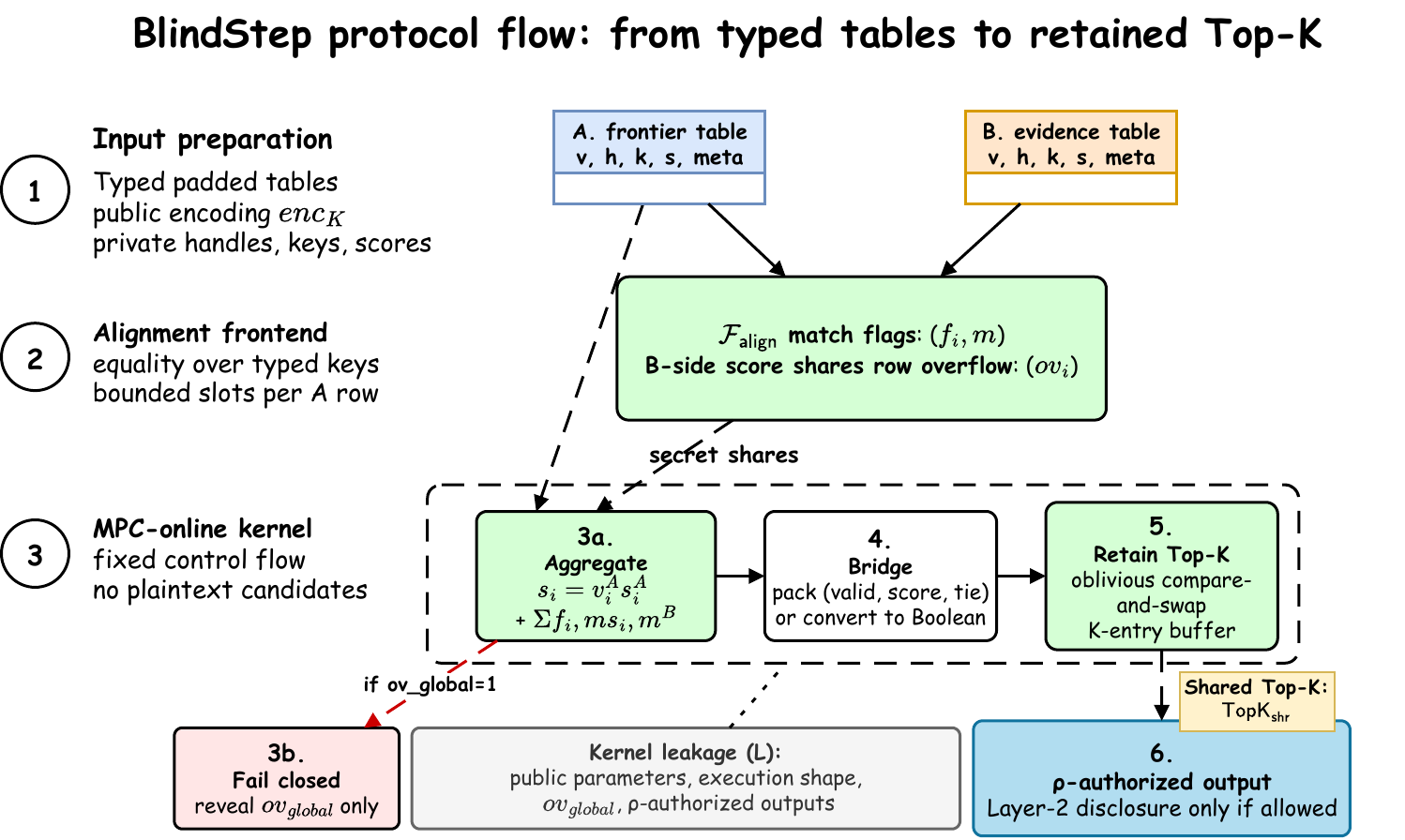}
    \caption{\BlindStep{} protocol flow. External alignment supplies bounded shares, the kernel performs aggregation, overflow checking, and secret-shared top-$K$ retention, and external policy $\rho$ governs reconstruction.}
    \label{fig:blindstep_protocol_flow}
\vspace{-3mm}
\end{figure}

Figure~\ref{fig:blindstep_protocol_flow} depicts the protocol flow around the kernel, complementing the system boundary in Figure~\ref{fig:blindstep_system_boundary}.
The MPC-online phase begins only after the external alignment frontend has applied the public compatibility rule and produced fixed-shape bounded slots.
The kernel therefore consumes Agent~$A$'s frontier-side shares and bounded-slot outputs, not raw keys or pairwise match records.

The kernel has three public-shape phases.
Module~1 ingests bounded slots, performs flag-weighted aggregation, checks the global overflow condition, and constructs a padded candidate table.
If the slot budget is violated, $\globaloverflow$ is revealed and the protocol aborts without releasing retained entries.
Otherwise, the representation bridge derives packed ranking values from the candidate table, H-realizations compare these values in the arithmetic domain, while B-realizations first convert them to Boolean shares.
Module~2 then performs secure retained-frontier top-$K$ selection and returns the secret-shared retained set $\SharedTopK$.

\subsection{Module~1: Bounded Candidate Ingestion and Aggregation}
\label{subsec:module1}

Module~1 turns the fixed-shape post-alignment interface into a padded candidate table.
It combines frontier-side shares supplied directly by Agent~$A$ with bounded-slot outputs produced by $\Falign{}$, then performs flag-weighted score aggregation and fail-closed overflow handling.
The MPC-online phase does not perform an $O(N^2)$ pairwise join, compatibility computation is handled by the external alignment frontend.
\smallskip

\noindent\textbf{Per-row inputs.}
For each frontier row $i$, Module~1 receives two types of secret-shared inputs.
First, Agent~$A$ supplies the frontier-side shares $\share{\valid_i^{(A)}}$ and $\share{\score_i^{(A)}}$.
Second, $\Falign{}$ supplies the fixed-shape bounded-slot output
\[
\left(
  \{(\share{f_{i,m}},\share{s^{(B)}_{i,m}})\}_{m=0}^{\Mmax-1},
  \share{\overflow_i}
\right).
\]
Each slot contains a match flag $f_{i,m}\in\{0,1\}$ and an extension-side score contribution $s^{(B)}_{i,m}\in[0,2^{b_{\mathsf{in}}})$, padding slots carry $f_{i,m}=0$.
The row-overflow bit $\overflow_i$ is supplied by $\Falign{}$ as a secret-shared indicator of slot-budget violation and is never opened individually.
\smallskip

\noindent\textbf{Flag-Weighted Aggregation.}
Module~1 aggregates the bounded slots for each frontier row as
\[
\score_i
=
\valid_i^{(A)}\cdot\score_i^{(A)}
+
\sum_{m=0}^{\Mmax-1}
  f_{i,m}\cdot s^{(B)}_{i,m}.
\]
All operations are evaluated over the corresponding secret shares in the MPC backend, the equation specifies the plaintext semantics of the computed share $\share{\score_i}$.
The factor $\valid_i^{(A)}$ suppresses frontier padding rows, and $f_{i,m}=0$ suppresses false matches and padding slots.
The loop structure, memory access pattern, and communication schedule depend only on public parameters such as $N$, $\Mmax$, and the configured score widths, not on the secret match pattern.
The aggregate score field uses $\scorebits$ bits.
To prevent aggregation overflow, $\scorebits$ must satisfy
\[
\scorebits \ge
\left\lceil
\log_2\!\left((\Mmax+1)(2^{b_{\mathsf{in}}}-1)+1\right)
\right\rceil .
\]
This bound ensures that the largest possible aggregated score fits in the $\scorebits$-bit score field.
\smallskip

\noindent\textbf{Fail-closed overflow semantics.}
Silently dropping excess matches beyond $\Mmax$ would alter the aggregated score and could shift the retained top-$K$ set.
Module~1 therefore audits the row-overflow indicators supplied by $\Falign{}$: it computes
\[
\globaloverflow=\bigvee_{i=0}^{N-1}\overflow_i
\]
and reveals only this single bit.
If $\globaloverflow=1$, the protocol follows the fail-closed abort path in Figure~\ref{fig:blindstep_protocol_flow}: no retained entries are released, and any intermediate shares produced before the abort are not reconstructed.
Otherwise, the candidate table continues to the representation bridge and Module~2.
\smallskip

\noindent\textbf{Candidate table construction.}
If no overflow is detected, Module~1 outputs a padded candidate table $\CandidateTable$ of size $N$.
Each row contains a candidate validity bit
\[
\valid_i =
\valid_i^{(A)}
\wedge
\mathbf{1}\!\left[
\sum_{m=0}^{\Mmax-1} f_{i,m} > 0
\right],
\]
the aggregate score $\score_i$, an opaque payload field, and a tie-breaking tag $\tie_i$ (Section~\ref{subsubsec:packed_representation}).
The threshold test is implemented as a secure comparison over secret-shared values and does not reveal the per-row match count.
Module~2 compares the ranking tuple $\tau_i=(\valid_i,\score_i,\tie_i)$.
The leading validity bit dominates the lexicographic order: any row with $\valid_i=0$ is ranked below every valid row, regardless of its aggregate score.
Thus a no-hit frontier row may have a non-zero intermediate aggregate score, but it cannot outrank any valid candidate.
The payload is carried through selection but is not used as a ranking key.
\smallskip

\noindent\textbf{Representation bridge.}
The representation bridge is a separate P2 phase between Module~1 and Module~2: Module~1 outputs a fixed-size arithmetic-shared candidate table, while Module~2 compares packed ranking values derived from $(\valid_i,\score_i,\tie_i)$, not compatibility keys or payload fields.
Depending on the realization, packed values may be compared in an arithmetic sharing domain or converted to Boolean shares before comparison.
In either case, the representation bridge is applied to all $N$ candidate rows, so its communication pattern and round schedule depend only on public parameters such as $N$, $b$, and $T$, not on the secret match pattern or candidate density.

\subsection{Module~2: Secure Top-$K$ Selection via Insertion Queue}
\label{subsec:module2}

Module~2 takes the padded candidate table from Module~1 and outputs the top-$K$ candidates under the lexicographic order on $(\valid_i,\score_i,\tie_i)$.
Since $K\ll N$, the target task is selection rather than full sorting.
\smallskip

\myparagraph{Packed Tuple Representation}
\label{subsubsec:packed_representation}
Each candidate row carries the validity bit, aggregate score, and tie-breaking tag produced by Module~1.
For comparison, we represent each candidate by the ranking tuple $\tau_i=(\valid_i,\score_i,\tie_i)$.

\myparagraphunderline{Tie tag}
The tie tag is internal ordering metadata used only to break exact ties under the lexicographic order $(\valid,\score,-\tie)$.
In the protocol definition, $\tie_i$ is sampled as an independent secret-shared value of width $T$, chosen so that collisions among $N$ rows are negligible (Appendix~\ref{app:tie_tag_width}).
The reveal policy $\policy$ does not authorize plaintext disclosure of tie tags.

\myparagraphunderline{Packed comparison}
When the tuple width fits within the backend's safe comparison width, the ranking tuple is packed into one integer. The packed layout from most significant to least significant bits is
\[
[\underbrace{\valid}_{1 \text{ bit}}]
[\underbrace{\score}_{\scorebits \text{ bits}}]
[\underbrace{\overline{\tie}}_{T \text{ bits}}],
\]
where $\overline{\tie}=(2^T-1)-\tie$ is the inverted tie tag.
Equivalently,
\[
\mathsf{pack}_{\scorebits,T}(\valid,\score,\tie)
=
\valid\cdot 2^{\scorebits+T}
+
\score\cdot 2^T
+
(2^T-1-\tie).
\]
For each ranking tuple $\EncTuple{i}=(\valid_i,\score_i,\tie_i)$, we write $\PackedTuple_i =
\mathsf{pack}_{\scorebits,T}(\valid_i,\score_i,\tie_i)$.
Integer comparison over packed values realizes the lexicographic order $(\valid\downarrow,\score\downarrow,\tie\uparrow)$: the leading validity bit ranks every valid tuple above every invalid tuple, among tuples with the same validity, larger aggregate scores rank higher, and exact ties are broken in favor of smaller $\tie$.
If $1+\scorebits+T$ exceeds the backend's safe comparison width, the same ordering is implemented by an unpacked lexicographic comparison over $(\valid,\score,\tie)$.

Our main full-pipeline configurations satisfy the packed-width bound:
with $N\le8192$, $\scorebits=16$, and
$T=\lceil 2\log_2 N\rceil+20$, the maximum width is
$1+16+46=63$ bits.
The selection buffer is initialized with the dummy sentinel
\[
\mathsf{pack}_{\scorebits,T}(\dummytuple)
=
\mathsf{pack}_{\scorebits,T}(0,0,0)
=
2^T-1 .
\]
This sentinel is dominated by any valid tuple, since $\valid=1$ sets the highest bit.
Tuples with $\valid=0$ remain as fillers only when fewer than $K$ valid candidates exist.
\smallskip

\myparagraph{Secure Insertion Queue Algorithm}
The secure insertion queue (Algorithm~\ref{alg:insertion-queue} in Appendix~\ref{app:correctness_proof}) maintains a size-$K$ buffer $B$ of dummy packed tuples.
For each candidate in the padded table, a chain of $K$ oblivious compare-and-swap (CAS) operations inserts it into the correct position if it belongs in the top-$K$.
Each buffer entry and carried value is secret-shared.
CAS uses secure packed-tuple comparison and an oblivious selector $\mathsf{Mux}(c,x,y)$, which returns $x$ when the secret bit $c=1$ and $y$ otherwise.
Both operations are data-oblivious: circuit execution and memory access depend only on public loop indices, not private tuple values.
\smallskip

\myparagraph{Complexity Analysis}
The insertion queue applies a fixed $K$-stage CAS chain to each of the $N$ padded candidates, and therefore performs exactly $NK$ oblivious compare-and-swap operations.
With packed ranking values of width $w=1+\scorebits+T$, let $C_{\mathsf{cmp}}(w)$ and $C_{\mathsf{swap}}(w)$ denote the backend-specific communication costs of one secure comparison and one conditional swap.

\begin{lemma}[Backend-Parametric Communication]
Module~2 incurs $O\!\left(NK\cdot
(C_{\mathsf{cmp}}(w)+C_{\mathsf{swap}}(w))\right)$ communication for insertion-queue top-$K$ selection.
\end{lemma}

\noindent
The bound follows from the fixed $NK$ CAS schedule, backend batching or mixed-domain lowering changes only the per-CAS costs, not the operation count or memory-access pattern.
In contrast, the Batcher full-ordering reference computes a complete order over all $N$ candidates and uses $\Theta(N\log^2N)$ comparisons.
The insertion queue exploits the retained-frontier regime $K\ll N$ and avoids this full-ordering cost.

\begin{theorem}[Correctness]
Upon termination, Algorithm~\ref{alg:insertion-queue} returns the $K$ candidate entries whose packed ranking tuples are largest under the lexicographic order.
\end{theorem}
\smallskip

\myparagraph{Correctness}
The proof follows from the standard insertion-sort buffer invariant: after processing the first $i$ candidates, the buffer contains the $K$ candidate entries with the largest packed ranking tuples among those $i$ inputs, dummy-padded when $i<K$.
We give the full induction proof in Appendix~\ref{app:correctness_proof}.
\smallskip

\myparagraph{Output Interface}
\label{subsec:output}
The Layer-1 output is the $K$-entry secret-shared retained set $\SharedTopK=\{\share{\mathsf{entry}_\ell}\}_{\ell=0}^{K-1}$.
Each retained entry carries the selected payload together with the secret-shared ranking fields at the kernel boundary shown in Figure~\ref{fig:blindstep_system_boundary}, any plaintext reconstruction is deferred to the external $\policy$-authorized disclosure step.
Layer-2 disclosure is governed by the reveal policy $\policy$ (Section~\ref{subsec:threat_model}), which excludes non-retained entries, local score contributions, and internal tie metadata.
The insertion queue is one realization of Module~2 within the \BlindStep{} interface, alternative selection backends are discussed in Section~\ref{sec:related}.

\section{Security Analysis}
\label{sec:security}

This section gives the simulation-based security analysis of \BlindStep{}.
We define the ideal functionality, refine the leakage function introduced in Section~\ref{subsec:threat_model}, state the security theorem, and sketch the simulator construction.
\subsection{Analysis Setup}
\label{subsec:security_model}

We analyze \BlindStep{} under the system model, adversarial model, and communication assumptions stated in Section~\ref{subsec:threat_model}.
The proof follows the real/ideal simulation paradigm: for each statically corrupted party $P\in\{A,B\}$, there exists a probabilistic polynomial-time simulator $\Sim_P$ whose ideal-world output is computationally indistinguishable from $P$'s real-world protocol view.
The analysis is per-invocation, composed leakage across repeated invocations is discussed in Section~\ref{subsec:repeated_invocation}.
\subsection{Ideal Functionality}
\label{subsec:functionality}

The functionality includes $\policy$-authorized Layer-2 disclosures as a modeling convention so that output leakage is explicit.
In the real workflow, as shown in Figure~\ref{fig:blindstep_system_boundary}, the kernel emits only Layer-1 secret shares and any authorized reconstruction is performed outside the kernel.
Thus the integrated formulation below is equivalent to a kernel-only execution with no plaintext opening composed with an external $\policy$-authorized reconstruction step.

We formalize \BlindStep{} in the $\Falign{}$-hybrid model.
The idealized alignment functionality $\Falign{}$ realizes the bounded candidate-composition interface: given the parties' private encoded keys in the public domain $\mathcal K$, the public compatibility rule $\compat$, and Agent~$B$'s extension-side score contributions, it outputs, for each frontier row $i$, a row-level overflow indicator and $\Mmax$ secret-shared bounded slots:
\[
\left(
\share{\overflow_i},
\left\{
\left(
\share{f_{i,m}},
\share{s^{(B)}_{i,m}}
\right)
\right\}_{m=0}^{\Mmax-1}
\right).
\]
Agent~$A$'s frontier-side validity bits and score contributions are provided directly to $\FBlindStep$, not as outputs of $\Falign{}$.
The ideal functionality $\FBlindStep$ operates on these bounded-slot outputs and does not receive raw extension rows, raw identifiers, or pairwise match patterns.

\begin{definition}[Ideal Functionality $\FBlindStep$]
\label{def:functionality}
Given public parameters $\pp=(N,K,\scorebits,\Mmax,\compat,\policy)$, frontier-side validity bits and score contributions, and the bounded-slot
outputs produced by $\Falign{}$, the functionality proceeds as follows.

\begin{enumerate}[label=(\arabic*), leftmargin=*, nosep]
    \item \textbf{Input.}
    For each frontier row $i\in\{0,\ldots,N-1\}$, receive the
    frontier-side validity bit $\valid_i^{(A)}$, score contribution
    $\score_i^{(A)}$, a row-level overflow indicator $\overflow_i$, and
    $\Mmax$ bounded slots. Each slot
    $m\in\{0,\ldots,\Mmax-1\}$ contains a match flag $f_{i,m}$ and an
    extension-side score contribution $s^{(B)}_{i,m}$.

    \item \textbf{Overflow check.}
    Compute $\globaloverflow=\bigvee_{i=0}^{N-1}\overflow_i$.
    If $\globaloverflow=1$, deliver $\globaloverflow$ to both parties,
    deliver no Layer-1 retained entries, and terminate with
    $\mathsf{abort}$.

    \item \textbf{Flag-weighted aggregation.}
    If $\globaloverflow=0$, compute
    \[
    \score_i
    =
    \valid_i^{(A)}\cdot\score_i^{(A)}
    +
    \sum_{m=0}^{\Mmax-1}f_{i,m}s^{(B)}_{i,m}.
    \]

    \item \textbf{Candidate-table construction.}
    Set
    \[
    \valid_i
    =
    \valid_i^{(A)}
    \wedge
    \mathbf{1}\!\left[
    \sum_{m=0}^{\Mmax-1} f_{i,m}>0
    \right].
    \]
    For each candidate row $i$, sample a fresh tie tag
    $\tie_i \leftarrow [0,2^T)$, where $T$ is the public tie-tag width
    defined in Section~\ref{subsubsec:packed_representation}.
    Construct the ranking tuple $(\valid_i,\score_i,\tie_i)$ together with
    the candidate payload. The tie tag is used only as internal ordering
    metadata for exact ties and is not disclosed by $\policy$.

    \item \textbf{Top-$K$ retention.}
    Select the $K$ retained entries under the lexicographic order
    $(\valid,\score,-\tie)$, and denote the resulting plaintext retained
    set by $\mathsf{TopK}$. Since $\valid$ is the leading field, any
    invalid row is ranked below every valid row regardless of its aggregate
    score.

    \item \textbf{Layer-1 output.}
    Secret-share the retained entries $\mathsf{TopK}$, including their payload and ranking fields, and deliver to each party its prescribed Layer-1 output share.

    \item \textbf{Layer-2 disclosure.}
    Apply the public reveal policy $\policy$ to the retained plaintext entries inside the ideal functionality.
    We write $(O_A^\policy,O_B^\policy)=\policy(\mathsf{TopK})$, where $O_A^\policy$ and $O_B^\policy$ are the Layer-2 disclosures authorized for Agent~$A$ and Agent~$B$, respectively.
    Deliver only $O_A^\policy$ to Agent~$A$ and only $O_B^\policy$ to Agent~$B$.
    This step models the external disclosure boundary rather than an additional plaintext output produced by the MPC-online kernel itself.
\end{enumerate}
\end{definition}

\subsection{Leakage Function}
\label{subsec:leakage}

The leakage function below captures the \BlindStep{} MPC-online kernel after $\Falign{}$ has produced fixed-shape bounded slots.
It therefore accounts for the kernel-leakage band shown in Figure~\ref{fig:blindstep_system_boundary}: public setup parameters, fixed execution-shape metadata, the fail-closed overflow bit, and $\policy$-authorized outputs.
Leakage from a concrete realization of $\Falign{}$ is accounted for separately and composes with this kernel leakage.

\begin{definition}[Kernel Leakage Function $\Leak$]
\label{def:leakage}
The kernel leakage function is $\Leak=(\LeakSetup,\LeakExec,\LeakAbort,\LeakOut)$.
The setup leakage is $\LeakSetup=(N,K,\scorebits,\Mmax,\compat,\policy)$.
The public description of $\compat$ includes the public key domain $\mathcal K$, the typed-key encoding format, and the output length.
For a fixed backend realization, the execution leakage is $\LeakExec=(\mathsf{prog},\mathsf{mem},\mathsf{rounds},\mathsf{msglen})$, where $\mathsf{prog}$ is the public program shape, $\mathsf{mem}$ is the memory-access schedule, $\mathsf{rounds}$ is the round count, and $\mathsf{msglen}$ is the vector of per-round message lengths.
These components are deterministic functions of the public parameters and the selected backend.
The abort leakage is $\LeakAbort=\globaloverflow$.
Finally, the output leakage is $\LeakOut=(O_A^\policy,O_B^\policy)$, where $(O_A^\policy,O_B^\policy)=\policy(\mathsf{TopK})$ are the Layer-2 disclosures authorized by the public reveal policy in the ideal functionality.
By default, $\policy$ does not disclose internal tie metadata or non-retained entries, such values are therefore not part of $\LeakOut$.
\end{definition}

The bit $\globaloverflow$ reveals only whether the public slot budget was satisfied: $\globaloverflow=1$ means that at least one row overflowed, and $\globaloverflow=0$ means that all rows stayed within budget.
It does not reveal which row overflowed, how many rows overflowed, or how many excess matches existed.

For a concrete deployment, the total leakage composes the chosen frontend's leakage with the kernel leakage above.
In our artifact, bounded-slot share files are produced by a VOLE-PSI/Circuit-PSI-derived frontend pipeline.
The frontend's security and leakage profile are external assumptions of the \BlindStep{} MPC-online analysis, while its costs and measurement scope are reported separately in Section~\ref{subsec:impl}.
\subsection{Security Statement}
\label{subsec:theorem}

The theorem is stated relative to the selected $\Falign{}$ realization: \BlindStep{} proves the MPC-online kernel after bounded slots have been produced, while frontend leakage, corruption assumptions, and input-consistency obligations are external and compose with $\Leak$.

\begin{theorem}[Security of \BlindStep{}]
\label{thm:security}
Assume that:
\begin{enumerate}[label=(\arabic*), leftmargin=*, nosep]
    \item The bounded candidate-composition interface securely realizes $\Falign{}$ over the public key domain $\mathcal K$, public compatibility rule $\compat$, and public slot budget $\Mmax$.
    \item The underlying MPC backend securely realizes the arithmetic and Boolean secret-sharing operations used for flag-weighted aggregation, secure comparison, and oblivious conditional selection.
    \item Any arithmetic-to-Boolean conversion used by a mixed-domain realization, e.g., edaBit-assisted conversion, is semi-honest secure.
    \item Tie tags are sampled independently of private inputs, row positions, local scores, and evidence order, and are not disclosed by the reveal policy $\policy$.
    \item Packed ranking tuples either fit within the backend's safe comparison width, or the protocol uses the unpacked lexicographic comparison fallback described in    Section~\ref{subsubsec:packed_representation}.
\end{enumerate}
Then $\PiBlindStep{}$ securely realizes $\FBlindStep$ with leakage $\Leak$ against static semi-honest adversaries corrupting at most one party.
\end{theorem}

\subsection{Proof Sketch}
\label{subsec:proof}

The proof follows standard sequential composition for semi-honest secure computation.
Module~1 operates on fixed-shape bounded-slot outputs of $\Falign{}$, the simulator uses the simulator for $\Falign{}$ and samples random shares for all intermediate flags, aggregate scores, validity bits, tie tags, packed ranking values, and payload fields, subject only to the corrupted party's input, $\globaloverflow$, and any $\policy$-authorized output.
Module~2 has a fixed $N\cdot K$ compare-and-swap structure over packed ranking tuples, the simulator reproduces the public execution shape and invokes the simulators for secure comparison, conditional selection, and any arithmetic-to-Boolean conversion.
Tie-tag shares are sampled as fresh shares of independent random values, matching the ideal functionality, and are not disclosed by $\policy$.
By sequential composition, the combined simulated view is indistinguishable from the real execution.
Appendix~\ref{app:security_proof} gives the full simulator.

\subsection{Leakage Implications}
\label{subsec:leakage_discussion}

Two leakage implications deserve mention.
First, if $\policy$ discloses retained-entry validity, observing fewer than $K$ valid retained candidates reveals that fewer than $K$ valid candidates existed.
This functionality-induced output leakage is captured by $\LeakOut$.
Second, $\globaloverflow$ is intentionally exposed instead of silently truncating evidence, allowing the protocol to enforce fail-closed correctness under the public slot budget, its information content is bounded by Definition~\ref{def:leakage}.

\subsection{Repeated Invocation and Composed Leakage}
\label{subsec:repeated_invocation}

The functionality and theorem are stated for a single invocation of \BlindStep{}.
For non-adaptive invocations with pre-committed public parameters $\{\pp^{(t)}\}_{t=1}^{T}$, the overall leakage is $\Leak^{1:T}=(\Leak^{(1)},\ldots,\Leak^{(T)})$, including the sequence of public parameters, fixed execution-shape metadata, global overflow bits, and any $\policy$-authorized outputs from each invocation.
This follows from sequential composition for semi-honest secure computation.

Repeated overflow bits or authorized outputs over correlated inputs may reveal coarse distributional information.
Experience-guided document reasoning~\cite{zai2026trace} illustrates why applications may retain state across steps.
Adaptive parameter selection, where $\pp^{(t+1)}$ depends on earlier outputs, requires modeling the orchestration controller, persistent cross-step state, and cross-invocation output policy, these are application-layer responsibilities outside the one-step interface.

\subsection{Scope and Limitations}
\label{app:limitations}

The security statement covers one semi-honest invocation of the
MPC-online kernel after bounded alignment. Malicious robustness,
adaptive security, frontend guarantees, reveal-policy enforcement, and
slot-budget validation are outside this analysis, see
Appendix~\ref{app:limitations}.
\section{Implementation and Evaluation}
\label{sec:evaluation}

We evaluate correctness, workload-grounded parameter choices, Module~2 scalability, integrated MPC-online performance, $\Mmax$ sensitivity, cost consistency across secret workload densities, and LAN/WAN behavior.
Our target deployments are batch-oriented or semi-interactive cross-organization analyses rather than latency-critical user-facing search.
Accordingly, the reported MPC-online costs should be interpreted within this setting.

\subsection{Implementation and Measurement Scope}
\label{subsec:impl}

We implement \BlindStep{} in MP-SPDZ~\cite{keller2020mp} using the
two-party semi-honest \texttt{semi2k} backend over
$\mathbb{Z}_{2^{64}}$, with MP-SPDZ's standard semi-honest
configuration and a 40-bit statistical security parameter.
Our artifact uses a VOLE-PSI/Circuit-PSI-derived frontend pipeline to
produce the fixed-shape bounded-slot share files consumed by the
MPC-online kernel.
We treat the frontend as an external component whose security and leakage profile depend on the selected $\Falign{}$ realization.
Therefore, we report its share-file generation costs separately from the main MPC-online measurements.

Integrated MPC-online experiments consume fixed-shape share files
containing secret shares of match flags, extension-side scores, and
row-overflow indicators. The frontend is not executed inline in the main
MPC-online measurements. Main tables report only the measured kernel
band in Figure~\ref{fig:blindstep_system_boundary}: Module~1 ingestion and
aggregation (P1), representation bridge or packing (P2), and Module~2
selection (P3). In breakdown tables, P1, P2, and P3 denote profiling
phases, not party identifiers. Artifact metadata, including commits,
compile flags, input-generation scripts, and \texttt{tc netem} commands,
is recorded in Appendix~\ref{app:artifact_metadata}.

\begin{table}[t]
\centering
\small
\setlength{\tabcolsep}{4pt}
\renewcommand{\arraystretch}{1.08}
\caption{Measurement scope of the main evaluation.}
\label{tab:measurement_scope}
\begin{tabularx}{\columnwidth}{@{}Xcc@{}}
\toprule
\textbf{Item} & \textbf{Included} & \textbf{Details} \\
\midrule
Module~1 ingestion/aggregation (P1) & Yes & Main tables \\
Bridge or packing (P2)              & Yes & Main tables \\
Module~2 selection (P3)             & Yes & Main tables \\
Frontend share-file generation      & No  & App.~\ref{app:e2e} \\
Offline preprocessing generation    & No  & App.~\ref{app:preproc} \\
Hardware and WAN emulation          & No  & App.~\ref{app:artifact_metadata} \\
\bottomrule
\end{tabularx}
\vspace{-3mm}
\end{table}
We evaluate six Module~2 realizations: \HQueue{}, \HSelect{}, \HSort{},
\BQueue{}, \BSelect{}, and \BSort{}. The H-variants compare packed
ranking values over arithmetic shares (\texttt{sint}) with
edaBit-assisted comparison lowering~\cite{escudero2020improved}, whereas
the B-variants convert packed ranking values to Boolean shares
(\texttt{sbitint}) before selection. In all variants, Module~1
aggregation is arithmetic. The three selection strategies are insertion
queue, $K$-pass repeated selection, and full sort using MP-SPDZ's
loop-based \texttt{Array.sort(batcher=True)}. The implementation invokes
MP-SPDZ's secure random-integer primitive once per candidate row,
producing one independent secret-shared tie tag per row, these shares are
used only for tie-breaking, never reconstructed, and excluded from
$\policy$. Communication and round counts are deterministic for fixed
compiled programs and public parameters, repeated executions affect only
wall-clock time.

\subsection{Correctness Verification}
\label{subsec:correctness}

Before reporting performance, we validate four properties: bounded one-to-many aggregation, fail-closed overflow handling, packed ranking-value injectivity, and cross-variant top-$K$ consistency.
All checks pass under controlled inputs: multi-slot injection covers no-, one-, two-, and three-hit rows, normal inputs with zero row-overflow shares complete, while controlled injection of a secret-shared overflow indicator triggers the fail-closed abort path, exhaustive round-trip over the $\scorebits=16$ score field reports zero collisions, and all six realizations return consistent top-$K$ retained sets on the fixed frontend share files used in Section~\ref{subsec:fullpipeline}.
These tests validate the MPC-online kernel behavior assuming the chosen $\Falign{}$ realization supplies correct bounded slots and row-overflow indicators, as required by Theorem~\ref{thm:security}.
Detailed test descriptions are in Appendix~\ref{app:correctness}.

\subsection{Evaluation Matrix}
\label{subsec:eval_matrix}

We evaluate \BlindStep{} with padded candidate-table sizes $N \in \{128,256,512,2048,8192\}$, retained frontier size $K=5$, and aggregate-score width $\scorebits=16$.
$\Mmax$ and WAN settings are evaluated separately in Sections~\ref{subsec:mmax} and~\ref{subsec:wan}.
We report MPC-online runtime, communication volume, and round count.
The six variants $\HSort{}, \HQueue{}, \HSelect{}, \BSort{}, \BQueue{}, \BSelect{}$ cross two sharing styles, hybrid arithmetic-Boolean and all-Boolean, with three top-$K$ algorithms: full sort, insertion queue,
and repeated selection.

\smallskip
\noindent\textbf{Comparison scope.}
Our full-sort variant is a full-ordering reference within
MP-SPDZ, not a state-of-the-art secure-selection baseline.
Fixing the backend, input shape, ranking semantics,
tie handling, and measurement scope isolates the cost
of full ordering relative to retained-frontier selection.
FSS-based protocols~\cite{agarwal2024secure} use
protocol-specific preprocessing and input interfaces,
while three-party radix sorting~\cite{asharov2022efficient}
assumes an honest majority.
Our experiments compare the six MP-SPDZ realizations.

\subsection{Real-World Workload Instantiation}
\label{subsec:workload}

To ground public parameters, we construct plaintext workloads from FB15k-237, ogbl-biokg, and ogbl-collab~\cite{toutanova2015fb15k, hu2020ogb}. For each dataset, we sample $M=1{,}000$ anchor nodes and compute one-hop candidate count $|A|$ and valid-survivor count $|J|$.
The long-tail candidate count motivates testing $N$ up to $8192$, and observed aggregate score magnitudes fit within the $\scorebits=16$ score field.
\begin{table}[t]
\centering
\small
\setlength{\tabcolsep}{4pt}
\renewcommand{\arraystretch}{1.08}
\caption{Workload statistics from $M=1{,}000$ anchors.}
\label{tab:workload_stats}
\begin{tabular}{@{}lrrrr@{}}
\toprule
\textbf{Dataset} & \textbf{P95 $|A|$} & \textbf{Max $|A|$} &
\textbf{Mean $|J|$} & \textbf{Score bits} \\
\midrule
FB15k-237   & 54  & 1{,}433  & 14.7 & 8 \\
ogbl-biokg  & 158 & 15{,}806 & 54.4 & 6 \\
ogbl-collab & 29  & 174      & 8.3  & 5 \\
\bottomrule
\end{tabular}
\vspace{-3mm}
\end{table}
At $N=512$, the padded table covers $98.9\%$-$100\%$ of sampled anchors, at $N=2048$, coverage exceeds $99.8\%$ (Appendix~\ref{app:workload_coverage}).
Anchors whose candidate count exceeds $N$ are not silently truncated: the fail-closed path requires either upstream filtering or a larger public $N$. $N=8192$ is a stress point grounded in the sampled workloads.

\smallskip

\noindent\textbf{Bounded-slot stress test.}
We measure per-candidate evidence multiplicity before invoking the MPC kernel to assess workload applicability.
Abort@$M$ is the fraction of sampled invocations in which at least one candidate's evidence multiplicity exceeds $M$.
At $\Mmax=10$, unfiltered abort rates are $90.4\%$, $99.3\%$, and $87.8\%$ on FB15k-237, ogbl-biokg, and ogbl-collab, respectively.
Per-relation filtering reduces FB15k-237's Abort@10 to $17.8\%$, and increasing $\Mmax$ to 50 reduces it to $1.1\%$.
However, ogbl-biokg retains Abort@100=$79.3\%$ after per-relation filtering, and ogbl-collab retains Abort@50=$37.9\%$ after top-5 pruning.
Deployments must therefore validate $\Mmax$ against post-filtering multiplicities and may require stronger task-specific pruning (Appendix~\ref{app:mmax_validation}).

\subsection{Selection Kernel Isolation}
\label{subsec:scalability}

\begin{figure}[t]
\centering
\begin{subfigure}[t]{0.49\linewidth}
    \centering
    \includegraphics[width=\linewidth]{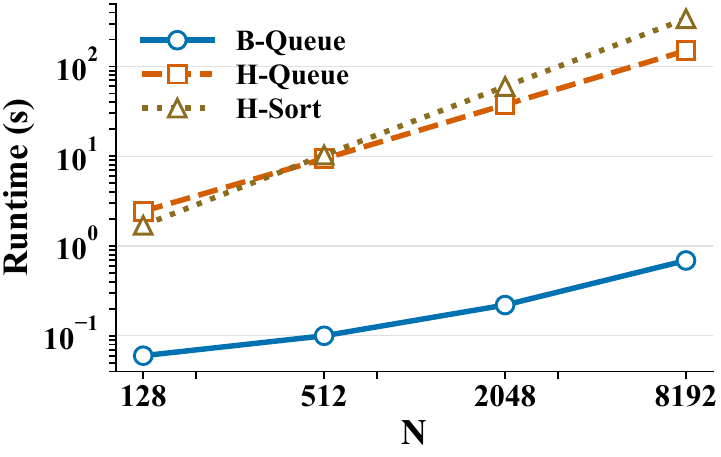}
    \caption{Runtime (s)}
    \label{fig:module2_scalability_n_runtime}
\end{subfigure}
\hfill
\begin{subfigure}[t]{0.49\linewidth}
    \centering
    \includegraphics[width=0.9\linewidth]{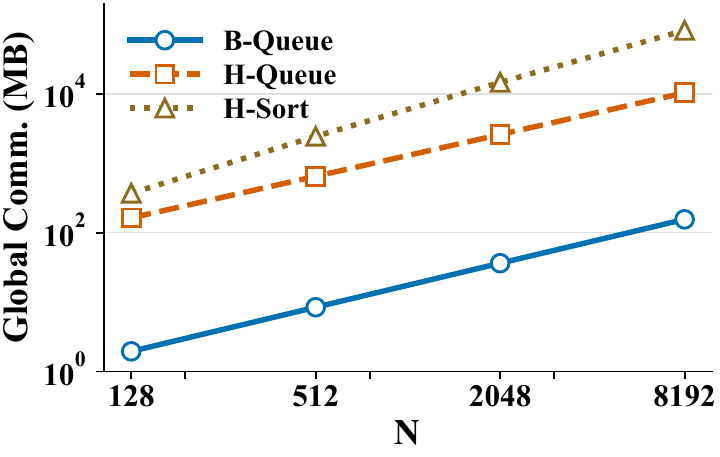}
    \caption{Global Comm. (MB)}
    \label{fig:module2_scalability_n_comm}
\end{subfigure}
\caption{Module~2-only scalability with synthetic packed ranking values
$(K=5,\scorebits=16,\text{LAN})$, shown on log-log scale.
}
\label{fig:module2_scalability_n}
\vspace{-3mm}
\end{figure}

To isolate Module~2's intrinsic scaling from Module~1 ingestion and representation bridging, we run the selection stage standalone with synthetic packed ranking values derived from $(\valid,\score,\tie)$, fixing $(K,\scorebits)=(5,16)$.
These inputs are pre-shared in each variant's native representation: \HQueue{} and \HSort{} consume arithmetic shares (\texttt{sint}), while \BQueue{} consumes Boolean shares (\texttt{sbitint}).
Bridge costs are therefore excluded from Figure~\ref{fig:module2_scalability_n} and included only in the integrated full-pipeline results (Table~\ref{tab:full_pipeline}).
Figure~\ref{fig:module2_scalability_n} reports runtime and communication
for these three representative configurations. \BQueue{} stays well below
one second across the evaluated range, while \HQueue{} and \HSort{} grow
at rates consistent with their $O(NK)$ and $\Theta(N\log^2N)$
comparison structures. At $N=8192$, \BQueue{}'s selection-kernel
communication is multiple orders of magnitude smaller than \HSort{}'s.
These standalone measurements complement the integrated results in
Section~\ref{subsec:fullpipeline}: the algorithmic gap between queue and
full sort is intrinsic to the selection stage, not an artifact of
Module~1 ingestion or bridge cost.

\subsection{MPC-Online Full Pipeline}
\label{subsec:fullpipeline}
We evaluate all six design variants on the integrated MPC-online pipeline using fixed-shape frontend share files across $N \in \{128, 256, 512, 2048, 8192\}$ with fixed $(K, b) = (5, 16)$.
The profiling phases P1/P2/P3 are defined in Section~\ref{subsec:impl}.
We use $N=256$ for the detailed per-phase breakdown.

\smallskip
\noindent\textbf{Full pipeline breakdown at $N=256$.}
Table~\ref{tab:full_pipeline} reports per-phase timings under LAN.
All variants share the same Module~1 semantics, differences arise from the backend representation and selection kernel after aggregation.
\begin{table}[t]
\centering
\small
\setlength{\tabcolsep}{3.5pt}
\renewcommand{\arraystretch}{1.08}
\caption{Full-pipeline MPC-online breakdown at $N=256$, $K=5$, $\scorebits=16$ under LAN.}
\label{tab:full_pipeline}
\begin{tabular}{@{}lrrrrrr@{}}
\toprule
\textbf{Var.} &
\textbf{P1} & \textbf{P2} & \textbf{P3} &
\textbf{Total} & \textbf{MB} & \textbf{Rounds} \\
\midrule
\HQueue{}  & 0.937 & $<$0.01 & 3.622 & 4.565 & 397.1     & 33{,}148 \\
\HSelect{} & 0.929 & $<$0.01 & 3.558 & 4.497 & 392.9     & 32{,}860 \\
\HSort{}   & 0.949 & $<$0.01 & 6.706 & 7.655 & 1{,}039.9 & 20{,}723 \\
\BQueue{}  & 0.284 & 0.104   & 0.323 & 0.720 & 10.7       & 35{,}587 \\
\BSelect{} & 0.285 & 0.104   & 0.327 & 0.720 & 11.1      & 35{,}338 \\
\BSort{}   & 0.290 & 0.104   & 0.584 & 0.968 & 15.6      & 45{,}351 \\
\bottomrule
\end{tabular}

\vspace{2pt}
\footnotesize
\parbox{\linewidth}{
{Note.} Results use fixed-shape frontend share files.
Times are in seconds, and communication is reported as global MB.
Values are medians over $n=7$ trials.
P1 is arithmetic aggregation, P2 is ranking-value packing for H-variants
or the arithmetic-to-Boolean bridge for B-variants, P3 is the Module~2
selection kernel.
Phase medians and total medians are computed separately, so rounded phase
values need not add exactly to the total.
}
\vspace{-3mm}
\end{table}
Within the hybrid domain, queue and repeated-select have comparable runtime, while full sort is slower and communicates roughly $2.6\times$ more (Table~\ref{tab:full_pipeline}).
All all-Boolean variants remain below one second, \BSort{} is the slowest B-variant due to its higher comparison count.
Across domains, all-Boolean variants are $6$-$8\times$ faster and use $38$-$70\times$ less communication than their hybrid counterparts under LAN.

\noindent\textbf{Scaling across $N$.}
Table~\ref{tab:multiscale} reports the total MPC-online runtime of the integrated pipeline as $N$ varies, with $n=5$ trials per cell ($n=10$ at $N=2048$, $n=7$ at $N=256$).
\begin{table}[t]
\centering
\small
\caption{Total MPC-online runtime (seconds, median) of the
integrated full pipeline at $(K,b)=(5,16)$, LAN. }
\label{tab:multiscale}
\begin{tabular}{@{}lrrrrr@{}}
\toprule
\textbf{Variant} & $N$=128 & $N$=256 & $N$=512 & $N$=2048 & $N$=8192 \\
\midrule
\HQueue{}  & 2.29 & 4.57 & 8.75  & 35.06   & 137.10 \\
\HSelect{} & 2.26 & 4.50 & 8.74  & 34.40   & 136.74 \\
\HSort{}   & 3.23 & 7.66 & 18.48 & 106.74  & 577.21 \\
\BQueue{}  & 0.38 & 0.72 & 1.28  & 5.00    & 18.86  \\
\BSelect{} & 0.40 & 0.72 & 1.30  & 4.96    & 19.30  \\
\BSort{}   & 0.50 & 0.97 & 2.20  & 10.97   & 57.43  \\
\bottomrule
\end{tabular}
\vspace{-3mm}
\end{table}
\HSort{} scales superlinearly with its $\Theta(N\log^2 N)$ Batcher structure, while queue and repeated-select variants scale near-linearly with the $O(N\cdot K)$ selection bound plus linear ingestion.
The \BQueue{}-vs-\HSort{} gap widens monotonically with $N$, reaching $30.4\times$ runtime and $290\times$ communication reduction at $N=8192$, the within-algorithm Boolean-vs-hybrid gap also strengthens with scale.
Detailed per-variant medians, trial spreads, communication volumes, and round counts are in Appendix~\ref{app:multiscale}.

\subsection{Sensitivity to Slot Budget}
\label{subsec:mmax}

\begin{figure}[t]
\centering
\begin{subfigure}[t]{0.49\linewidth}
    \centering
    \includegraphics[width=\linewidth]{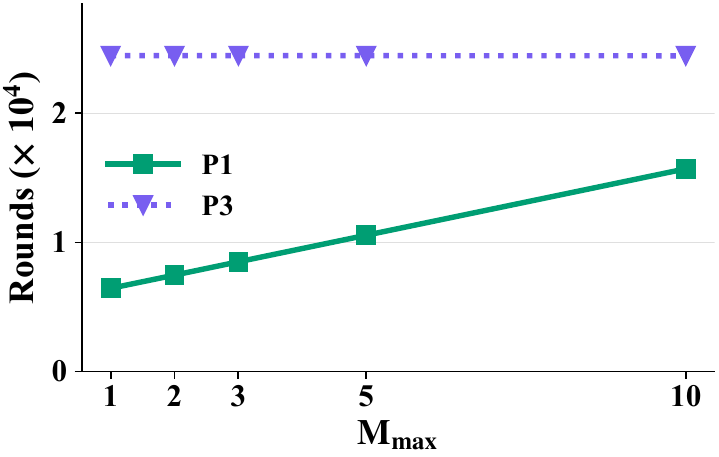}
    \caption{Round counts}
    \label{fig:mmax_sensitivity_rounds}
\end{subfigure}
\hfill
\begin{subfigure}[t]{0.49\linewidth}
    \centering
    \includegraphics[width=\linewidth]{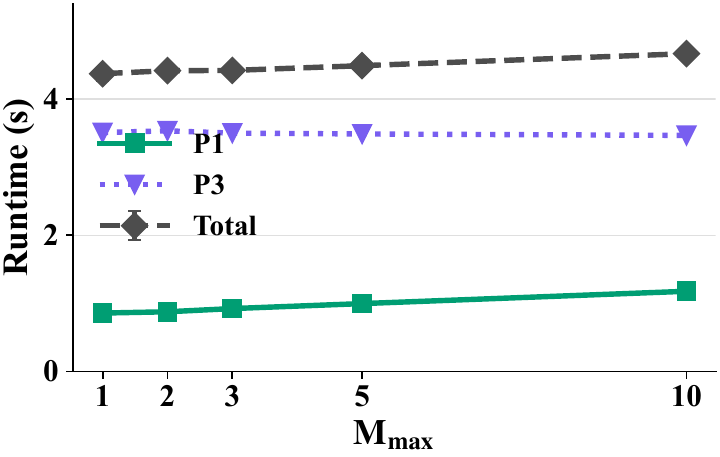}
    \caption{Median runtime}
    \label{fig:mmax_sensitivity_runtime}
\end{subfigure}
\caption{$\Mmax$ sensitivity at $N=256$, $K=5$, $\scorebits=16$ under LAN, for tested values $\{1,2,3,5,10\}$.
P1 aggregation rounds grow with the slot budget, while selection-phase rounds remain unchanged.
Runtime values are medians over three trials.}
\label{fig:mmax_sensitivity}
\vspace{-1mm}
\end{figure}

We evaluate $\Mmax$ using \HQueue{} at $(N,K,\scorebits)=(256,5,16)$.
Figure~\ref{fig:mmax_sensitivity} reports round counts and median runtime for $\Mmax\in\{1,2,3,5,10\}$.
Module~1 rounds increase from $6{,}462$ to $15{,}723$, consistent with the $O(N\Mmax)$ ingestion bound, while selection-phase rounds remain at $24{,}475$.
Because $\Mmax$ should be chosen from the post-filtering multiplicity distribution rather than from raw graph degree, the per-candidate-row $\text{P99}=45$ of per-relation-filtered FB15k-237 (Table~\ref{tab:mmax_validation}) motivates a larger budget.
A separate three-trial campaign extends the sweep to $\Mmax=50$: relative to its own $\Mmax=10$ baseline, median total runtime rises by $38.9\%$, global communication by $10.9\%$, and Module~1 rounds grow about $3.6\times$, while selection-phase rounds stay at $24{,}475$.
For input scores of at most eight bits, $\Mmax=50$ requires $\lceil\log_2(51\cdot255+1)\rceil=14$ aggregate bits, within $\scorebits=16$.
Appendix~\ref{app:mmax} reports the detailed measurements and the density sweep.

\subsection{Network Sensitivity (WAN)}
\label{subsec:wan}

We evaluate all six full-pipeline variants under WAN-emulated conditions (100\,ms RTT, 100\,Mbit/s, injected via \texttt{tc netem} on the loopback interface).
These runs use the same $N=256$, $K=5$, and $\scorebits=16$ setting as Table~\ref{tab:full_pipeline}, enabling direct LAN-WAN comparison.

\noindent\textbf{Full-pipeline WAN results.}
Table~\ref{tab:wan_full} reports median WAN runtimes over three trials.
Under WAN, all variants are round-dominated: local computation and communication volume become secondary to per-round latency.
\HSort{} achieves the lowest WAN total ($1{,}028$\,s) because it has the smallest round count among the six variants, despite its higher communication.
The all-Boolean variants, which are fastest under LAN, lose their advantage under high latency because their larger round counts dominate.

\begin{table}[t]
\centering
\small
\caption{MPC-online full-pipeline WAN results using fixed-shape frontend share files ($N=256$, $K=5$, $\scorebits=16$, 100\,ms RTT, 100\,Mbit/s).}
\label{tab:wan_full}
\begin{tabular}{@{}lrrrrrr@{}}
\toprule
\textbf{Variant} &
\textbf{P1} & \textbf{P2} & \textbf{P3} &
\textbf{Total} & \textbf{Comm.} & \textbf{Rounds} \\
\midrule
\HQueue{}  & 428 &   $<$1 & 1{,}226 & 1{,}654 &   397.1 & 33{,}148 \\
\HSelect{}  & 428 &   $<$1 & 1{,}212 & 1{,}640 &   392.9 & 32{,}860 \\
\HSort{}  & 428 &   $<$1 &   600   & 1{,}028 & 1{,}039.9 & 20{,}723 \\
\BQueue{}  & 422 & 334 & 1{,}029 & 1{,}786 &     10.7 & 35{,}587 \\
\BSelect{}  & 422 & 334 & 1{,}016 & 1{,}773 &    11.1 & 35{,}338 \\
\BSort{}    & 421 & 334 & 1{,}519 & 2{,}275 &    15.6 & 45{,}351 \\
\bottomrule
\end{tabular}

\vspace{2pt}
\footnotesize
\parbox{\linewidth}{
{Note.} Times are in seconds, communication is global MB, values are medians over $n=3$ trials.
Measurement scope follows Table~\ref{tab:measurement_scope}. P1/P2/P3 use the same profiling-phase definitions as Table~\ref{tab:full_pipeline}.
Under WAN, runtime is dominated by round count.}
\vspace{-4mm}
\end{table}

\smallskip
\noindent\textbf{RTT sweep.}
An additional RTT sweep for the two representative variants \BQueue{} and \HSort{} is reported in Appendix~\ref{app:wan_rtt}.
Across 25\,ms, 50\,ms, and 100\,ms RTT at $N=256$, \HSort{} remains faster despite over $100\times$ higher communication, with its advantage growing from $1.38\times$ to $1.74\times$.
This supports the interpretation that round latency can dominate communication volume in the evaluated WAN regime, without implying a universal ordering for larger $N$ or other bandwidth settings.

\noindent\textbf{Large-$N$ WAN stress comparison.}
Appendix~\ref{app:wan_n2048_stress} reports a two-variant large-$N$ WAN stress check rather than a full WAN sweep.
In this evaluated high-latency setting, \HSort{} remains $1.14\times$ faster than \BQueue{} despite about $200\times$ higher communication, reinforcing that round count can remain the binding cost.

\begin{figure}[t]
\centering
\begin{subfigure}[t]{0.49\linewidth}
    \centering
    \includegraphics[width=\linewidth]{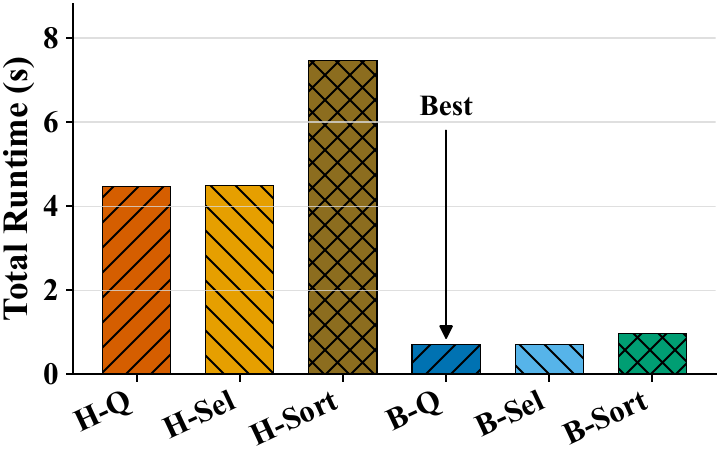}
    \caption{LAN}
    \label{fig:lan_total_runtime}
\end{subfigure}
\hfill
\begin{subfigure}[t]{0.49\linewidth}
    \centering
    \includegraphics[width=0.9\linewidth]{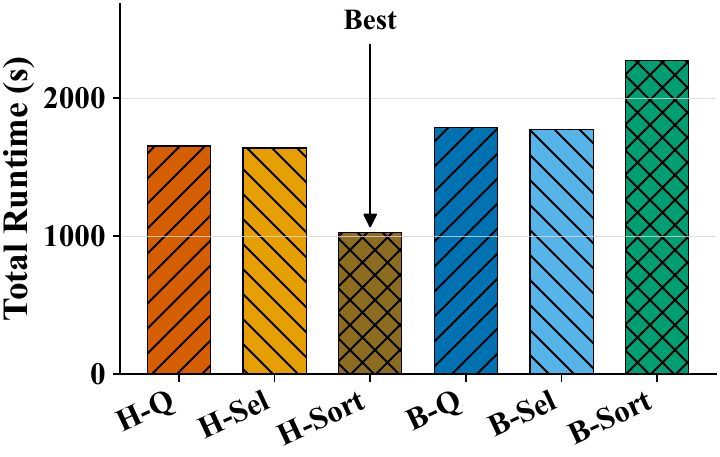}
    \caption{WAN}
    \label{fig:wan_total_runtime}
\end{subfigure}
\caption{MPC-online total runtime under LAN and WAN ($N=256$, $K=5$, $\scorebits=16$), y-axis scales differ.
The fastest realization shifts from \BQueue{}/\BSelect{} under LAN to \HSort{} under the evaluated WAN profile.}
\vspace{-5mm}

\label{fig:lan_wan_runtime}
\end{figure}

\noindent\textbf{LAN vs. WAN optimality shift.}
Figure~\ref{fig:lan_wan_runtime} summarizes the total-runtime change between LAN and WAN.
Under LAN, \BQueue{} and \BSelect{} are jointly fastest ($0.720$\,s), under WAN, \HSort{} becomes fastest ($1{,}028$\,s).
Thus, the LAN-optimal realization is not WAN-optimal: under LAN, runtime tracks local computation and communication, whereas under the evaluated WAN profile it tracks round count.

\noindent\textbf{Deployment takeaway and evaluation scope.}
The preferred realization depends on the network regime and dominant pipeline stage.
These conclusions apply to the MPC-online phase under the evaluated configurations: main tables exclude frontend execution and offline preprocessing, as summarized in Table~\ref{tab:measurement_scope}.
The six-variant WAN table uses one main emulated profile, and the RTT sweep covers only two representative variants. These experiments expose round-latency sensitivity but do not map the full crossover surface across larger $N$, bandwidths, or geographic deployments.

\textit{
Appendices~\ref{app:tie_tag_width}--\ref{app:limitations} cover tie-tag sizing, packed-value safety, the insertion-queue algorithm and correctness proof, the detailed simulator proof, and scope limitations.
Appendices~\ref{app:impl_measurement_details}--\ref{app:workload} report implementation and measurement details, frontend and preprocessing accounting, correctness and robustness checks, and workload instantiation.
Appendix~\ref{app:additional_experiments} provides additional experiments, including large-scale full-pipeline results, $\Mmax$ sensitivity, Module~1 density-obliviousness checks, WAN RTT sensitivity, and a large-$N$ WAN stress check.
}
\section{Conclusion}
\vspace{-2mm}
\label{sec:conclusion}

In this paper, we
present \BlindStep{}, a fixed-shape MPC-online kernel for a recurring post-alignment step in cross-party reasoning pipelines.
By introducing a bounded-slot interface for task-filtered evidence, fail-closed overflow handling, top-$K$ frontier retention, and an explicit per-invocation leakage boundary, \BlindStep{} provides a small interface for composing and retaining cross-party evidence after private alignment.
Our MP-SPDZ realizations show that backend choice is bottleneck-dependent rather than universal.
In the evaluated semi2k 2PC setting, low-communication Boolean variants perform best under LAN, while the lower-round \HSort{} reference is fastest under the $N=256$, $100$\,ms RTT WAN profile.
These results provide matched-scope evidence for backend selection, while also showing that larger $N$, bandwidth-constrained networks, frontend and offline costs, and stronger security models may shift the trade-offs.
Future work includes matched cross-framework selection benchmarks, malicious-secure hardening, and leakage accounting for adaptive multi-step workflows, especially when repeated invocations create cumulative disclosure risks.

\vspace{2mm}
\noindent
\textbf{Acknowledgments.}
Xiaoyang Wang is supported by the ARC grants DP230101445 and DP240101322.

\clearpage
\begingroup
\def\IEEEbibitemsep{0.15\baselineskip plus 0.05\baselineskip minus 0.05\baselineskip}

\endgroup

\clearpage
\appendices

\section{Tie-Tag Width and Packed-Value Safety}
\label{app:tie_tag_width}
\vspace{-3mm}

We set $T=\lceil 2\log_2 N\rceil+\sigma$, where $\sigma$ is a statistical slack parameter.
The tie domain is $[0,2^T)$, so the birthday bound gives collision probability at most $\frac{N(N-1)}{2^{T+1}} \le 2^{-\sigma-1}$.
Our artifact uses $\sigma=20$.
The implementation enforces the packed-width check $1+b+T\le 63$, leaving one spare bit in the 64-bit ring for safe unsigned comparison.
Configurations exceeding this bound use the unpacked lexicographic comparison fallback described in Section~\ref{subsubsec:packed_representation}.
Tie tags are sampled by independent invocations of the backend's secure random primitive.
The row index does not influence the sample.

\section{Insertion-Queue Algorithm and Correctness}
\label{app:correctness_proof}
\vspace{-3mm}

\begin{algorithm}[!ht]
\small
\caption{Secure Top-$K$ Selection via Insertion Queue}
\label{alg:insertion-queue}

\KwIn{Padded candidate table $\{\PackedTuple_i\}_{i=0}^{N-1}$ of
secret-shared packed tuples; public parameters $(N,K,\scorebits,T)$}
\KwOut{Top-$K$ packed tuples in descending lexicographic order}
\tcp{$\mathsf{Mux}(c,x,y)$ returns $x$ if secret bit $c=1$, else $y$}

\tcp{Initialize the buffer with dummy packed tuples}
\For{$j \gets 0$ \KwTo $K-1$}{
    $B[j] \gets \mathsf{pack}_{b,T}(0,0,0)$\;
}

\For{$i \gets 0$ \KwTo $N-1$}{
    $x \gets \PackedTuple_i$\;
    \For{$j \gets 0$ \KwTo $K-1$}{
        $\mathsf{old} \gets B[j]$\;
        $\mathsf{gt} \gets (x > \mathsf{old})$\;
        \tcp{Secure comparison over packed values}

        $B[j] \gets \mathsf{Mux}(\mathsf{gt},x,\mathsf{old})$\;
        $x \gets \mathsf{Mux}(\mathsf{gt},\mathsf{old},x)$\;
        \tcp{Oblivious conditional selection}
    }
}

\Return{$B[0], \ldots, B[K-1]$}\;
\end{algorithm}

\noindent\textbf{Correctness proof.}
We prove the standard insertion-buffer invariant.
After processing $i$ candidates, $B$ contains the $K$ largest packed ranking values among those candidates, dummy-padded when $i<K$.
The claim is trivial for $i=0$.
For candidate $i$, the $K$-stage CAS chain compares the carried value $x$ with each buffer position and keeps the larger value in the buffer while carrying the smaller value forward.
Thus an entering top-$K$ value is inserted and lower-ranked values cascade downward, otherwise the candidate exits the buffer.
The invariant holds for $i+1$, and at $i=N$ the buffer contains the top-$K$ packed ranking values.
Payload fields, when present, are swapped with their associated ranking values and do not affect comparison.

\section{Detailed Security Proof}
\label{app:security_proof}
\vspace{-2mm}

We prove Theorem~\ref{thm:security} via a modular composition argument.
Fix a corrupted party $P\in\{A,B\}$.
The simulator $\Sim_P$ receives $P$'s prescribed input, the public parameters, the leakage $\Leak$, and the Layer-2 disclosure $O_P^\policy$ authorized for $P$ by the reveal policy $\policy$.
It does not receive the honest party's raw input, non-retained candidates, pairwise match patterns, or local score contributions.

\myparagraph{C.1. Module~1: Bounded Candidate Ingestion and Aggregation}
Module~1 operates only on the fixed-shape secret-shared bounded-slot outputs of $\Falign{}$. For corrupted party $P$, the simulator $\Sim_P$ invokes the simulator guaranteed for $\Falign{}$ using $P$'s prescribed input and the leakage allowed by $\Leak$.
Given the public parameters, $\globaloverflow$, and $O_P^\policy$, $\Sim_P$ simulates the Module~1 transcript as follows.
If $\globaloverflow=1$, it simulates the same public abort point and no Layer-1 retained entries.
If $\globaloverflow=0$, it samples uniformly random shares for match flags, extension-side score contributions, candidate validity bits, aggregate scores, tie tags, packed ranking values, and payload fields, subject only to the corrupted party's input shares and to consistency constraints imposed by the ideal output delivered to that party.
Since a single party's view of additive or Boolean secret shares is uniformly distributed in the real execution, these simulated shares are indistinguishable from real shares.
The loop structure, memory-access pattern, message lengths, and round schedule are fixed by $(N,\Mmax,\scorebits)$ and therefore simulatable from $\LeakExec$.

\myparagraph{C.2. Module~2: Secure Top-$K$ Selection}
Module~2 performs $N \cdot K$ oblivious compare-and-swap operations.

\myparagraphunderline{Data-Oblivious Execution}
The control flow of Algorithm~\ref{alg:insertion-queue} depends only on public parameters: the outer loop iterates exactly $N$ times, the inner loop iterates exactly $K$ times, and memory access patterns are fixed.
No control-flow branches or memory accesses depend on secret values.

\myparagraphunderline{Comparison and Multiplexer Security}
Each comparison $x>B[j]$ and conditional selection $\mathsf{Mux}(\mathsf{gt},\cdot,\cdot)$ operates on secret-shared values.
By the security of the underlying primitives, these operations reveal neither the compared values nor the secret comparison result $\mathsf{gt}$.
The simulator does not need to know the packed ranking values or comparison outcomes, it only reproduces the fixed execution shape and the authorized final output.

\myparagraphunderline{Tie-tag simulation}
The simulator samples tie-tag shares as fresh shares of independent random values, matching the ideal functionality and the sampling rule in Appendix~\ref{app:tie_tag_width}.
Since tie tags are independent of private inputs and are not disclosed by $\policy$, the simulated view does not require the honest party's input row order.
The retained Layer-1 shares, $\globaloverflow$, and any $\policy$-authorized Layer-2 disclosure are simulated from the ideal output and leakage.

\myparagraphunderline{Simulator Construction}
The simulator $\Sim_P$ proceeds as follows:
(1) Receive $P$'s prescribed input, public parameters, leakage $\Leak$, and the Layer-2 disclosure $O_P^\policy$ authorized for $P$.
(2) Simulate the bounded candidate-composition interface using the simulator guaranteed for $\Falign{}$.
(3) Simulate Module~1 by sampling random shares and messages consistent with $P$'s own input, $\globaloverflow$, and $O_P^\policy$.
(4) Simulate Module~2 by invoking the simulators for secure comparison, arithmetic-to-Boolean conversion when used, and oblivious conditional selection.
(5) Return exactly the Layer-2 plaintext disclosures authorized for $P$ by $\policy$, and no other plaintext values.
\noindent
Because the control flow, memory-access patterns, message lengths, and round counts depend only on public parameters, and all intermediate candidate values remain secret-shared, $\Sim_P$ requires only the corrupted party's prescribed input/output and the leakage $\Leak$ to produce an indistinguishable view.

\myparagraph{C.3. Composition}
The security of the composed protocol follows from standard sequential composition for semi-honest secure computation~\cite{oded2009foundations,lindell2017simulate}.
Module~1 produces secret-shared candidate entries, including payload and ranking fields, consumed by Module~2.
By the sequential composition theorem for semi-honest secure computation, replacing each idealized component with its secure realization preserves indistinguishability of the corrupted party's view.
Therefore, the composed protocol realizes $\FBlindStep$ with $\Leak$.

\section{Scope and Limitations}
\label{app:limitations}
\vspace{-2mm}

\noindent\textbf{Threat model.}
Our guarantee holds against static semi-honest adversaries.
Malicious robustness, fairness, input-consistency enforcement, adaptive security, and implementation side channels are outside the present theorem.
The deterministic abort captured by $\LeakAbort$ is not a defense against maliciously induced selective aborts.

\noindent\textbf{External components.}
The alignment frontend and reveal-policy enforcement are external assumptions of the \BlindStep{} theorem. Our artifact uses equality over typed encoded identifiers, richer predicates require separate frontend leakage analysis.

\noindent\textbf{Bounded slot budget.}
The public bound $\Mmax$ is an interface contract for typed, application-filtered inputs, not a universal bound on raw graph degree or evidence multiplicity.
Violations trigger fail-closed abort rather than silent truncation.
Selecting and validating $\Mmax$ is deployment-specific and evaluated through workload analysis rather than by the protocol itself.

\noindent\textbf{Functionality scope.}
\BlindStep{} is limited to additive flag-weighted aggregation followed by hidden-score top-$K$ retention.
Supporting capped sums, nonlinear scores, or learned ranking functions would require a different functionality and separate leakage analysis.

\section{Implementation and Measurement Details}
\label{app:impl_measurement_details}
\vspace{-2mm}

\myparagraph{E.1. Artifact and Reproducibility Metadata}
\label{app:artifact_metadata}
The artifact records the MP-SPDZ commit, the VOLE-PSI Circuit-PSI frontend commit, compile commands, frontend share-file generation scripts, WAN-emulation commands, and raw run logs used for the reported tables.
MPC programs use MP-SPDZ \texttt{semi2k} over $\mathbb{Z}_{2^{64}}$, integrated MPC-online runs consume fixed-shape frontend share files, while Module~2-only runs use synthetic packed ranking values.
Experiments run on a single Linux server with an Intel Xeon Gold 6248R CPU and 384\,GB RAM, with both parties co-located on localhost.
The main WAN profile uses \texttt{tc netem} with 50\,ms one-way delay and a 100\,Mbit/s cap, emulating approximately 100\,ms RTT, additional RTT-sensitivity runs are reported in Appendix~\ref{app:wan_rtt}.
Wall-clock experiments use $n=3$ trials by default, main LAN full-pipeline tables (Section~\ref{subsec:fullpipeline}, Appendix~\ref{app:multiscale}) use $n=7$ or more trials per variant, communication and round count are deterministic for a fixed compiled program and public parameter setting.

\myparagraph{E.2. Frontend Share-File Generation}
\label{app:e2e}
Table~\ref{tab:e2e_lan} reports artifact frontend share-file generation time together with representative MPC-online costs. This measurement captures the VOLE-PSI/Circuit-PSI-derived input-generation pipeline used to produce fixed-shape share files, it is not a full distributed PSI end-to-end network measurement. Share-file generation remains below \(0.5\)~s across all evaluated sizes and falls below \(10\%\) of \BQueue{} end-to-end time for \(N\ge512\). Adding this cost does not change the LAN ordering of the reported MPC-online variants.

\begin{table}[!ht]
\centering
\small
\caption{LAN accounting with artifact frontend share-file generation.}
\label{tab:e2e_lan}
\begin{tabular}{@{}lrrrr@{}}
\toprule
\textbf{Variant} & $\mathbf{N}$ & \textbf{Frontend} & \textbf{MPC online} & \textbf{Frontend \%} \\
\midrule
\BQueue{} & 128  & 0.126 & 0.38  & 24.9 \\
\HSort{}  & 128  & 0.126 & 3.23  & 3.8  \\
\BQueue{} & 256  & 0.145 & 0.72  & 16.8 \\
\HSort{}  & 256  & 0.145 & 7.66  & 1.9  \\
\BQueue{} & 512  & 0.129 & 1.28  & 9.2  \\
\HSort{}  & 512  & 0.129 & 18.48 & 0.7  \\
\BQueue{} & 2048 & 0.194 & 5.00  & 3.7  \\
\HSort{}  & 2048 & 0.194 & 106.74 & 0.2 \\
\BQueue{} & 8192 & 0.418 & 18.86 & 2.2 \\
\HSort{}  & 8192 & 0.418 & 577.21 & 0.1 \\
\bottomrule
\end{tabular}

\smallskip
\footnotesize

\parbox{\linewidth}{
{Note:} Frontend and MPC-online times are in seconds.
Frontend time is the median wall-clock time of frontend share-file generation over three trials.
MPC-online time is the median from Table~\ref{tab:multiscale}.
Frontend \% is computed as $\mathrm{Frontend}/(\mathrm{Frontend}+\mathrm{MPC\ online})$.}
\vspace{-2mm}

\end{table}

\begin{table}[h]
\centering
\small
\caption{Compile-time preprocessing material estimates at
$(N,K,\scorebits,\Mmax)=(256,5,16,3)$. }
\label{tab:preproc_estimates}
\begin{tabular}{@{}lrrr@{}}
\toprule
\textbf{Variant} & \textbf{Int. triples} &
\textbf{edaBits} & \textbf{Bit triples} \\
\midrule
\HQueue{}  & 188{,}222       & ---             & ---             \\
\HSelect{} & $\infty^\ast$   & ---             & ---             \\
\HSort{}   & 494{,}022       & ---             & ---             \\
\BQueue{}  & 2{,}304         & 513             & 405{,}110       \\
\BSelect{} & 2{,}304         & 513             & $\infty^\ast$   \\
\BSort{}   & 2{,}304         & 513             & 811{,}959       \\
\bottomrule
\end{tabular}

\smallskip
\footnotesize
\parbox{\linewidth}{
{Note.} ``---'' indicates that the static analyzer reports no material of that type for the variant. $^{\ast}$For the Select variants, MP-SPDZ's static analyzer reports an unbounded estimate for the material consumed by the selection stage because of its compiler-level loop structure: integer triples for H-Select, which compares over arithmetic shares, and bit triples for B-Select, which compares over Boolean shares. B-Select's integer-triple and edaBit counts arise in the fixed-shape aggregation and conversion stages and are bounded. This affects only the static preprocessing estimate, not the measured online runtime, communication, or round counts reported in the main tables.}
\vspace{-2mm}

\end{table}

\myparagraph{E.3. Preprocessing Material Estimates}
\label{app:preproc}
Table~\ref{tab:preproc_estimates} reports MP-SPDZ static-analyzer preprocessing estimates at \((N,K,b,\Mmax)=(256,5,16,3)\). These are compile-time upper bounds, the toolchain does not expose exact runtime consumption of preprocessing material.
These counts indicate relative preprocessing-material demand within MP-SPDZ, not end-to-end deployment cost.

\section{Correctness and Robustness Details}
\label{app:correctness}
\vspace{-2mm}

\myparagraph{F.1. One-to-Many Aggregation}
\label{app:correctness:onetomany}
We test the Module~1 aggregation from Section~\ref{subsec:module1} on controlled multi-slot inputs with \(N=256\) and \(\Mmax=3\). Slot \(m=0\) follows the frontend share-file format, controlled non-zero values are additionally injected into slot \(m=1\) for row \(A[3]\) and
slots \(m=1,2\) for row \(A[6]\), covering no-, one-, two-, and three-hit cases. The test reconstructs the intermediate aggregate score \(\score_i\) and compares it with the expected value (Table~\ref{tab:app_onetomany}).
\begin{table}[!ht]
\centering
\small
\caption{Detailed one-to-many aggregate-score check
$(N=256,\Mmax=3)$.}
\label{tab:app_onetomany}
\begin{tabular}{@{}lccc c@{}}
\toprule
\textbf{Row} & \textbf{Case} & \textbf{Slots} &
\textbf{Expected} & \textbf{Actual} \\
\midrule
$A[1]$ & no-hit    & 0 & 6   & -   \\
$A[3]$ & two-hit   & 2 & 321 & 321 \\
$A[6]$ & three-hit & 3 & 657 & 657 \\
$A[9]$ & one-hit   & 1 & 139 & 139 \\
\bottomrule
\end{tabular}
\vspace{0.3mm}
\begin{flushleft}
\footnotesize
Note: Expected and actual denote the aggregate score before and after secure aggregation.
\end{flushleft}
\vspace{-3mm}
\end{table}
For the no-hit row A[1], the expected aggregate is the frontier-side contribution alone ($\score_1 = 6$), since no extension-side slot contributes.
The check reports aggregates only for rows with $\valid_i = 1$, so no actual value is listed. With $\valid_1 = 0$, the packed ranking tuple is ordered below every valid candidate regardless of the intermediate aggregate score, so the row cannot displace a valid candidate from the retained frontier.

\myparagraph{F.2. Overflow Fail-Closed Check}
\label{app:correctness:overflow}
The \(\Falign{}\) interface supplies \(M_{\max}\) bounded slots and a secret-shared row-overflow indicator \(ov_i\) for each row. This check does not validate how a deployment-specific frontend detects multiplicity overflow, it validates the MPC-online kernel's fail-closed response once such indicators are provided. In the normal run, all row-overflow shares are zero and the protocol completes. In the injected run, one row is assigned \(ov_i=1\) using \texttt{BLINDSTEP\_INJECT\_OVERFLOW}, the kernel reveals only \(\globaloverflow\) and aborts before releasing retained entries. This confirms that out-of-contract state is treated as a protocol failure rather than silently absorbed into aggregation.

\myparagraph{F.3. Packed Ranking-Value Injectivity}
\label{app:correctness:injectivity}
Module~2 compares packed ranking values derived from $\tau_i=(\valid_i,\score_i,\tie_i)$.
The implementation packs each ranking tuple as
\[
\mathsf{pack}_{\scorebits,T}(\valid,\score,\tie)
=
\valid\cdot 2^{\scorebits+T}
+
\score\cdot 2^T
+
(2^T-1-\tie).
\]
Under the bounds $\valid<2$, $\score<2^{\scorebits}$, and $\tie<2^T$, the fields occupy disjoint bit ranges.
Hence the encoding is injective whenever $1+\scorebits+T$ fits within the backend's safe comparison width.
The leading validity bit ensures that every valid tuple dominates every invalid tuple, regardless of the invalid tuple's aggregate score.
For the default correctness-test configuration, we use $(N,\scorebits)=(256,16)$ and $T=8$. We exhaustively test $\valid\in\{0,1\}, \score\in[0,2^{16}),\tie\in[0,256)$, covering $33{,}554{,}432$ ranking tuples.
The test performs a pack-and-unpack round trip and verifies that no two distinct tuples map to the same packed value, it reports zero collisions.

\myparagraph{F.4. Cross-Variant Retained-Set Consistency}
\label{app:correctness:crossvariant}
Using the fixed frontend share files from Section~\ref{subsec:fullpipeline},
we run all six realizations and compare the reconstructed retained sets
authorized for testing. All variants return the same top-\(K\) retained
set and order, confirming that representation bridging and backend
choice do not change the retained-frontier semantics.

\section{Workload Instantiation Details}
\label{app:workload}
\vspace{-2mm}

\myparagraph{G.1. Padded Candidate-Table Coverage}
\label{app:workload_coverage}
Table~\ref{tab:app_n_coverage} reports the fraction of sampled anchors covered by each padded candidate-table size, complementing Table~\ref{tab:workload_stats}.
The results show that \(N=512\) already covers nearly all sampled anchors, except for a small long-tail fraction in ogbl-biokg, and \(N=2048\) covers at least \(99.8\%\) across all three workloads.
Anchors outside the public padded size are not

\begin{table}[!ht]
\centering
\small
\caption{Coverage of sampled anchors at each padded candidate-table size $N$.}
\label{tab:app_n_coverage}
\begin{tabular}{@{}lrrrr@{}}
\toprule
\textbf{Dataset} & $N=128$ & $N=512$ & $N=2048$ & $N=8192$ \\
\midrule
FB15k-237   & 99.4\% & 99.9\% & 100.0\% & 100.0\% \\
ogbl-biokg  & 93.2\% & 98.9\% & 99.8\%  & 99.9\%  \\
ogbl-collab & 99.7\% & 100.0\% & 100.0\% & 100.0\% \\
\bottomrule
\end{tabular}
\end{table}

\begin{table}[!ht]
\centering
\small
\setlength{\tabcolsep}{3pt}
\caption{Bounded-slot contract stress test under raw and simple-filtered
workloads.}
\label{tab:mmax_validation}
\begin{tabular}{@{}lrrrrrr@{}}
\toprule
 & \multicolumn{2}{c}{FB15k-237} & \multicolumn{2}{c}{ogbl-biokg}
   & \multicolumn{2}{c}{ogbl-collab} \\
\cmidrule(lr){2-3} \cmidrule(lr){4-5} \cmidrule(lr){6-7}
 & raw & per-rel & raw & per-rel & raw & top-5 \\
\midrule
\multicolumn{7}{@{}l}{\emph{Evidence multiplicity, per candidate row}} \\
P95         & 440  & 16   & 13531 & 6113  & 95   & 93   \\
P99         & 1433 & 45   & 25810 & 25810 & 196  & 184  \\
\addlinespace
\multicolumn{7}{@{}l}{\emph{Abort rate per invocation (\%)}} \\
$\Mmax=3$   & 95.7 & 29.2 & 99.8  & 98.7  & 98.7 & 98.6 \\
$\Mmax=10$  & 90.4 & 17.8 & 99.3  & 94.6  & 87.8 & 86.5 \\
$\Mmax=50$  & 79.2 & 1.1  & 97.5  & 84.4  & 44.7 & 37.9 \\
$\Mmax=100$ & 69.5 & 0.2  & 95.1  & 79.3  & 19.6 & 13.5 \\
\bottomrule
\end{tabular}
\vspace{0.3mm}

\begin{flushleft}
\footnotesize
Note: an invocation counts as an abort if \emph{any} of its candidate rows
carries evidence multiplicity exceeding the public slot budget, the
denominator is the 1{,}000 sampled anchors per dataset used in
Section~\ref{subsec:workload}.
\end{flushleft}
\end{table}

\noindent
silently truncated, deployments must either choose a larger public \(N\), apply task-specific filtering, or route the instance to the fail-closed path described in Section~\ref{subsec:module1}.

\myparagraph{G.2. Bounded-Slot Contract Stress Test}
\label{app:mmax_validation}
This stress test is a deployment diagnostic: it measures how often evidence multiplicity would exceed the public slot budget before invoking the MPC kernel.
It is not a kernel correctness test and should not be interpreted as the intended filtered workload.
If the budget is exceeded, the protocol's correct response is the fail-closed abort path, since silently truncating excess evidence could change aggregate scores and the retained top-\(K\) set.
Table~\ref{tab:mmax_validation} reports raw multiplicities and simple filtering checks.
The results show that simple generic filters can reduce tail multiplicity in some structured typed workloads, but are not universally sufficient.
Deployments whose workloads remain above the public slot budget must apply stronger task-specific pruning, split high-multiplicity typed rows into multiple public instances, choose a larger \(\Mmax\) subject to score and packed-width constraints, or use a different aggregation functionality.
Quantifying the semantic effect of such filtering, such as retrieval recall or task accuracy, is application-dependent and outside this kernel-level evaluation.

\section{Additional Experiment and Analysis}
\vspace{-2mm}
\label{app:additional_experiments}

This appendix collects supporting measurements that are too detailed for the main evaluation but help interpret backend choice: large-scale LAN accounting, slot-budget sensitivity, fixed-shape cost checks, and WAN round-latency stress tests.

\myparagraph{H.1. Multi-Scale Full-Pipeline Details}
\label{app:multiscale}
Table~\ref{tab:app_multiscale_selected} reports the largest integrated MPC-online setting used in Section~\ref{subsec:fullpipeline}.
At \(N=8192\), \BQueue{} is \(30.6\times\) faster than \HSort{} and communicates \(290\times\) less data, \BSelect{} is nearly tied with \BQueue{}, while \BSort{} remains faster than all H-variants but pays the expected full-ordering cost.

\begin{table}[t]
\centering
\small
\caption{Integrated MPC-online accounting at \(N=8192,K=5,b=16\) under
LAN. Values are medians over \(n=5\) trials.}
\label{tab:app_multiscale_selected}
\begin{tabular}{@{}lrrr@{}}
\toprule
\textbf{Variant} & \textbf{Total (s)} &
\textbf{Spread (s)} & \textbf{Comm. (MB)} \\
\midrule
\HQueue{}  & 137.10 &  0.90 & 12{,}620 \\
\HSelect{} & 136.74 &  2.56 & 12{,}616 \\
\HSort{}   & 577.21 & 13.45 & 84{,}799 \\
\BQueue{}  &  18.86 &  0.70 &    316 \\
\BSelect{} &  19.30 &  0.85 &    317 \\
\BSort{}   &  57.43 &  0.77 &    944 \\
\bottomrule
\end{tabular}

\vspace{2pt}
\footnotesize
\parbox{\linewidth}{
\emph{Note.} Spread is max-minus-min wall-clock time over repeated
trials, communication is global MB and deterministic for each compiled
program and public parameter setting.}
\vspace{-1mm}

\end{table}

\smallskip

\begin{table}[t]
\centering
\small
\caption{$\Mmax$ sensitivity details ($N=256$, $K=5$, $\scorebits=16$, LAN).}
\label{tab:app_mmax_sensitivity}
\begin{tabular}{@{}lrrrrr@{}}
\toprule
$\Mmax$ & \textbf{P1} & \textbf{P1 Rds.} &
\textbf{P3} & \textbf{Total} & \textbf{Spread} \\
\midrule
1  & 0.858 &  6{,}462 & 3.509 & 4.373 & 0.205 \\
2  & 0.877 &  7{,}495 & 3.534 & 4.417 & 0.025 \\
3  & 0.924 &  8{,}519 & 3.500 & 4.421 & 0.139 \\
5  & 0.997 & 10{,}576 & 3.488 & 4.491 & 0.088 \\
10 & 1.178 & 15{,}723 & 3.467 & 4.667 & 0.069 \\
\bottomrule
\end{tabular}

\vspace{2pt}
\footnotesize
\parbox{\linewidth}{
{Note.} Times are in seconds and reported as medians over $3$ trials.
Spread denotes the max-min range of total runtime over $3$ trials.
P3 rounds are 24{,}475 for every evaluated $\Mmax$.}
\vspace{-3mm}

\end{table}

\begin{table}[!ht]
\centering
\small
\caption{Module~1 cost under varying match densities
($N=256$, $\Mmax=3$, LAN, single-run timings).}
\label{tab:app_dist_oblivious}
\begin{tabular}{@{}lrrrr@{}}
\toprule
$\delta$ & \textbf{Hits} & \textbf{P1} & \textbf{Comm.} & \textbf{Rounds} \\
\midrule
0\%  & 0   & 0.286 & 69.4 & 9{,}036 \\
10\% & 26  & 0.292 & 69.4 & 9{,}036 \\
50\% & 128 & 0.277 & 69.4 & 9{,}036 \\
90\% & 231 & 0.285 & 69.4 & 9{,}036 \\
\bottomrule
\end{tabular}

\vspace{2pt}
\footnotesize
\parbox{\linewidth}{
Note. Times are in seconds and communication is global MB.
P1 denotes the profiling phase covering share loading and Module~1 flag-weighted aggregation. P1 timings re-measured after the input-alignment change under a lighter-load environment, not comparable in magnitude to Table 4.}
\end{table}

\myparagraph{H.2. $\Mmax$ Sensitivity Details}
\label{app:mmax}
Table~\ref{tab:app_mmax_sensitivity} reports timing, round-count, and run-to-run variability values underlying Figure~\ref{fig:mmax_sensitivity}.
As $\Mmax$ grows from 1 to 10, Module~1 rounds increase \(2.43\times\), but median total runtime increases only \(6.7\%\) because selection dominates this \HQueue{} configuration.
This supports the main claim that $\Mmax$ is primarily an ingestion-side cost knob once the packed score width is fixed.
A same-batch supplementary measurement at \((N,K,\scorebits)=(256,5,16)\) extends the budget to \(\Mmax=50\) (three runs, medians).
Relative to its own \(\Mmax=10\) baseline in that batch, raising \(\Mmax\) to 50 increases total runtime by \(38.9\%\) and global communication by \(10.9\%\), while Module~1 rounds grow about $3.6\times$ and selection-phase rounds remain at \(24{,}475\).
Module~1 rounds thus overtake the selection-phase count between \(\Mmax=10\) and \(\Mmax=50\), consistent with the larger runtime increase observed over that range.

\myparagraph{H.3. Module~1 Density Sweep}
\label{app:density_sweep}
Table~\ref{tab:app_dist_oblivious} reports the Module~1 density sweep summarized in Section~\ref{subsec:mmax}.
The experiment varies the PSI match density while keeping the public parameters fixed.
Communication and round count remain identical across all settings, and P1 runtime varies by less than \(2\%\) from the median.
This supports the fixed-shape claim for Module~1: online cost is determined by public parameters rather than the secret match density.
The experiment uses the isolated \texttt{blindstep\_module1} program and does not instantiate a Module~2 selection variant.

\begin{table}[!ht]
\centering
\small
\setlength{\tabcolsep}{4pt}
\caption{RTT sensitivity for two representative variants at
$N=256,K=5,b=16$. Times are medians over three trials. The 100\,ms rows are the corresponding entries from Table~\ref{tab:wan_full}.}
\label{tab:wan_rtt_sweep}
\begin{tabular}{@{}lrrrr@{}}
\toprule
\textbf{RTT} & \textbf{Variant} & \textbf{Total (s)} & \textbf{Comm. (MB)} & \textbf{Rounds} \\
\midrule
25\,ms  & \BQueue{} & 451.5  & 10.7      & 35{,}587 \\
25\,ms  & \HSort{}  & 326.3  & 1{,}039.9 & 20{,}723 \\
50\,ms  & \BQueue{} & 896.5  & 10.7      & 35{,}587 \\
50\,ms  & \HSort{}  & 558.8  & 1{,}039.9 & 20{,}723 \\
100\,ms & \BQueue{} & 1{,}786 & 10.7      & 35{,}587 \\
100\,ms & \HSort{}  & 1{,}028 & 1{,}039.9 & 20{,}723 \\
\bottomrule
\end{tabular}
\end{table}
\begin{table}[!ht]
\centering
\footnotesize
\setlength{\tabcolsep}{3pt}
\caption{Large-\(N\) WAN stress check under
\((N,K,b)=(2048,5,16)\), 100\,ms RTT, and 100\,Mbit/s bandwidth.}
\label{tab:wan_n2048_stress}
\begin{tabular}{@{}lrrrrrr@{}}
\toprule
\textbf{Variant}
& \textbf{P1}
& \textbf{P2}
& \textbf{P3}
& \textbf{Total}
& \textbf{Comm.}
& \textbf{Rounds} \\
\midrule
\BQueue{} & 3{,}312 & 2{,}688 & 8{,}227 & 14{,}229 & 81.5   & 283{,}861 \\
\HSort{}  & 3{,}409 & 1.7     & 9{,}118 & 12{,}529 & 15{,}264.6 & 251{,}329 \\
\bottomrule
\end{tabular}

\vspace{2pt}
\footnotesize
\parbox{\linewidth}{
\emph{Note.} Times are in seconds, communication is global MB, and
rounds are deterministic for each compiled program and public parameter
setting. This is a two-variant stress check, not a full large-\(N\) WAN
sweep.}

\end{table}
\myparagraph{H.4. WAN RTT Sensitivity}
\label{app:wan_rtt}
Table~\ref{tab:wan_rtt_sweep} separates the two representative variants across 25--100\,ms RTT while keeping bandwidth fixed.
The lower-round \HSort{} reference is \(1.38\times\), \(1.60\times\), and \(1.74\times\) faster than \BQueue{} at 25, 50, and
100\,ms RTT, respectively, despite communicating over \(100\times\) more data.
This confirms that the 100\,ms WAN reversal is not a single-profile artifact within the evaluated \(N=256\) setting.
We use the sweep only to support the round-latency interpretation, it is not a full WAN crossover surface.

\smallskip
\myparagraph{H.5. Large-$N$ WAN Stress Check}
\label{app:wan_n2048_stress}
This two-variant stress check complements the \(N=256\) WAN profile in Section~\ref{subsec:wan}, it tests whether the round-latency effect persists at a larger candidate-table size rather than characterizing the full WAN crossover surface. Result is shown in Table~\ref{tab:wan_n2048_stress}.
\HSort{} remains \(1.14\times\) faster at \(N=2048\) even though it communicates about \(200\times\) more data.
The result is driven by fewer rounds and by the WAN cost of the \BQueue{} representation bridge, it should be read as a stress check for the evaluated high-latency profile, not as a universal preference for full sorting.

\vspace{-2mm}
\section*{Ethics and LLM Usage Statement}
\vspace{-2mm}
\myparagraph{Ethical Considerations}
This work proposes a privacy-preserving computation primitive and does not involve human subjects, private user-data collection, or deployment against third-party systems.
Workload studies use public datasets only to derive benchmark parameters.
Deployment requires application-level governance over the reveal policy $\policy$, upstream preparation, and authorized disclosures.

\smallskip
\myparagraph{LLM Usage}
Large language models were used to assist with language polishing, organization, and non-authoritative editing suggestions.
The authors are responsible for all technical content, experimental results, security claims, proofs, citations, and final text.

\end{document}